\documentclass[12pt]{article}
\usepackage{amsfonts}
\usepackage{amssymb}
\usepackage{cite}
\def\hybrid{\topmargin -20pt    \oddsidemargin 0pt
        \headheight 0pt \headsep 0pt
        \textwidth 6.25in       
        \textheight 9.5in       
        \marginparwidth .875in
        \parskip 5pt plus 1pt   \jot = 1.5ex}

\hybrid
\newcommand{\beqn}{\begin{eqnarray}}
\newcommand{\eeqn}{\end{eqnarray}}
\newcommand{\be}{\begin{equation}}
\newcommand{\ee}{\end{equation}}
\newcommand{\non}{\nonumber \\}
\newcommand{\vol}{{\cal V}}
\newcommand{\bi}{\bar{\imath}}
\newcommand{\bj}{\bar{\jmath}}
\newcommand{\bI}{\bar{I}}

\newcommand{\pu}{\partial_{\mu}}
\newcommand{\po}{\partial^{\mu}}
\newcommand{\fn}{\footnotesize}
\newcommand{\str}{\mbox{\fn str}}
\newcommand{\tm}{\tilde{M}}

\newcommand{\tvol}{\tilde{\vol}}
\newcommand{\tg}{\tilde{G}}
\newcommand{\ba}{\bar{\alpha}}
\newcommand{\bz}{\bar{Z}}
\newcommand{\bb}{\bar{b}}
\newcommand{\bk}{\bar{k}}
\newcommand{\bm}{\bar{m}}

\newcommand{\oneone}{e}

\def\R{({\rm Re}f)}
\def\I{({\rm Im}f)}

\def\rr{r}
\def\rs{r_s}
\def\bb{b}

\def\Gi{\hat{G}}
\def\nN{\hat{N}}

\begin {document}
\begin{titlepage}
\begin{center}

\hfill hep-th/9912181\\
\vskip 2cm
{\large \bf Duality in  Heterotic Vacua 
With Four Supercharges}\footnote{Work 
supported by:
GIF -- the German--Israeli
Foundation for Scientific Research,
DAAD -- the German Academic Exchange Service
and the Landesgraduiertenf\"orderung
Sachsen-Anhalt.}

\vskip .5in

{\bf Michael Haack\footnote{email: {\tt michael@hera1.physik.uni-halle.de}}
\  and Jan Louis\footnote{email: {\tt j.louis@physik.uni-halle.de}}}\\

\vskip 0.8cm
{\em Fachbereich Physik, Martin-Luther-Universi\"at Halle-Wittenberg,\\
Friedemann-Bach-Platz 6, D-06099 Halle, Germany}

\end{center}

\vskip 2.5cm

\begin{center} {\bf ABSTRACT } \end{center}

We study heterotic vacua with four supercharges
in three and four space-time dimensions
and their duals obtained as M/F-theory 
compactified on Calabi-Yau fourfolds.
We focus on their respective moduli spaces and derive the 
K\"ahler potential  for heterotic vacua obtained as circle compactifications
of four-dimensional $N=1$ heterotic theories.
The K\"ahler potential of the dual theory is computed by compactifying
11-dimensional supergravity on  
Calabi-Yau fourfolds.
The duality between these theories is checked
for K3-fibred fourfolds and an appropriate 
F-theory limit is discussed.
 
\vskip 1cm

\vfill

December 1999
\end{titlepage}

\section{Introduction} 
\label{sectintro}

The $E_8\times E_8$ heterotic string has been
the prime candidate for providing
a string theoretic version of the (supersymmetric)
Standard Model for over a decade.
In particular, the vacuum solutions with 4 Minkowskian
space-time dimensions ($D=4$) and $N=1$ supersymmetry have been
extensively studied due to their phenomenological prospects.
However, despite considerable efforts 
a number of (serious) problems remain within the framework
of the perturbative heterotic string such as 
a missing mechanism for hierarchical supersymmetry
breaking and the stabilization of the dilaton.

With the advent of string dualities two things have changed.
On the one hand it has been possible to control some
of the non-perturbative properties (or couplings) 
of string theory \cite{dualrev}. On the other hand the
$SO(32)$ heterotic string and even more so the type I string
have been investigated from the point of view of
viable particle phenomenology \cite{tI}.

In this paper we continue to focus on 
$N=1$ vacua of the heterotic string but include 
their circle compactifications to
$D=3$ into our considerations.
The reason is that in $D=3$
the vector multiplet 
contains a real scalar in the adjoint representation
of the heterotic gauge group and thus  
-- contrary to the situation in $D=4$ --
a Coulomb branch exists.
Some of the non-perturbative
features of these vacua are believed to be captured by
F-theory compactified on elliptic Calabi-Yau fourfolds
$Y_4$  or M-theory compactified on $Y_4$ for the three-dimensional
case \cite{V,MV,MV2,KL,BJPS,FMW,CD,AC,AC2,AC3,WNPW,DGW,W2,CL,L,ACL,
GC,BA,PM,BM,PM2,KLRY,KS,GM,M,R,BS,BLS,BLS2,LSW,GVW,DRS,KV,
BJPSV,DM,DR,BCGJL,SVW,BB}.

The purpose of this paper is to facilitate a comparison 
of the corresponding effective Lagrangians 
and in particular the  
identification of those M/F-theory vacua which have a 
perturbative heterotic limit.
In our analysis we ignore all
charged matter multiplets and 
only focus on the moduli which are constrained by
supersymmetry to parameterize a 
K\"ahler manifold \cite{Zumino,DNT}. 
We derive the K\"ahler potential for the moduli of 
a Calabi-Yau fourfold   in the large 
volume limit of M-theory 
including the moduli coming from the 
three-form potential and compare it to 
the K\"ahler potential of the 
corresponding heterotic vacua.
In spirit our analysis is very close to 
a similar analysis carried out
for the duality of type IIA on a Calabi-Yau threefold 
$Y_3$ and the 
heterotic string on K3$\times T^2$ in 
refs.\ \cite{KLM,AL}.

The organisation of the paper is as follows. In section 
\ref{secthet3d} (and appendix~B) 
we start with a general effective 
supergravity Lagrangian in $N=1, D=4$ which can be obtained 
by compactifications of the heterotic string on an
appropriate  
(0,2) superconformal field theory\footnote{This includes
in particular compactifications on Calabi-Yau 
threefolds $Y_3$ with a choice of the gauge bundle.} 
and perform a Kaluza-Klein 
reduction to $D=3$ on a circle. 
We derive the three-dimensional K\"ahler potential in terms 
of the four-dimensional one and an additional term coming 
from the four-dimensional vector fields. In section 
\ref{sectm} (and appendix~C)
we compactify 11-dimensional supergravity on a 
Calabi-Yau fourfold with vanishing
Euler number ($\chi=0$) and vanishing
four-form flux.
We derive the K\"ahler 
potential for the moduli
which include the 
geometrical moduli arising as deformations
of the 
K\"ahler class and complex structure 
of the fourfold as well as the moduli
coming from the expansion of the 
M-theory three-form.
The M-theory K\"ahler potential is  
compared to the heterotic K\"ahler potential
in section 
\ref{duality}. We  show that for
fourfolds which are K3-fibrations over 
a complex two-dimensional base $B_2$
the two K\"ahler potentials agree in 
the limit of large $B_2$ and 
weak heterotic coupling.
Finally, the F-theory limit is discussed.

Section~5 contains our conclusions
and appendix~A assembles notation and  conventions.
A preliminary version of some of our results was given in \cite{HL}. 

\section{Heterotic String Vacua}
\label{secthet3d}

\subsection{$D=4, N=1$ heterotic vacua}
\label{secthet3dein}

Our starting point is a generic
effective supergravity Lagrangian in $D=4$ with 
$N=1$ supersymmetry.
Such vacua  can be constructed as compactifications 
of the heterotic string on Calabi-Yau threefolds
$Y_3$
or more generally from appropriate $(0,2)$ superconformal field theories.
For the purpose of this paper it is sufficient
to focus only on the vector multiplets and the chiral moduli multiplets and 
ignore all charged matter multiplets.
The effective Lagrangian in this case reads \cite{FL}\footnote{We neglect
the possibility of anomalous $U(1)$ gauge factors with appropriate
four-dimensional Green-Schwarz terms. 
}
\beqn
{\cal{L}}^{(4)} &=& \frac{1}{2} R^{(4)} - 
G_{\bar IJ}^{(4)} (\Phi,\bar{\Phi}) \partial_m
\bar{\Phi}^{\bar I} \partial^m \Phi^{J} 
- \frac{1}{4} \mbox{Re}  f_{ab}(\Phi) 
F^{a}_{mn} F^{b mn} 
+ 
\frac{1}{4} \mbox{Im} f_{ab}(\Phi) F^{a}_{mn} 
\tilde F^{b mn} \nonumber \\
&&+ \ldots \quad ,
\label{eqhet4d}
\eeqn
where  $m,n=0,\ldots,3$,
$\Phi^I$ are the moduli fields
and $F^{a}_{mn}$ is the field strength of the gauge bosons
$A_m^a$. The index $a$ labels the generators of the gauge group
$G$ and thus $a=1,\ldots, \mbox{dim}(G)$.  
$N=1$ supersymmetry requires that the $f_{a b}(\Phi)$ are holomorphic functions
of the moduli which are 
 further constrained by gauge invariance.
The metric $G_{\bar{I}J}^{(4)}$ has to be a K\"ahler metric,
that is 
\be
G_{\bar{I}J}^{(4)} = \bar\partial_{\bar{I}}\partial_J
K^{(4)}(\Phi,\bar{\Phi})\ ,
\ee
where $K^{(4)}$ is the K\"ahler potential.

Perturbative heterotic string theory imposes further constraints
on the functions $K^{(4)}$ and  $f_{a b}$.
First of all the rank r of the gauge group $G$ is bounded by the 
central charge $c$ of the left moving (bosonic) conformal field theory 
on the worldsheet.
For heterotic strings in $D=4$ one has $c=22$ and hence
\be 
{\rm r}(G) \leq 22. 
\label{rGbound}
\ee
Secondly, the holomorphic $f_{a b}$ are universal at
the string tree level and determined by the heterotic dilaton $\lambda_4$
and the axion $a$ which is the (Poincar\'e)
dual of the antisymmetric 
tensor $B_{mn}$.
The two scalars are combined into a complex 
$S=\lambda_4^{-2} + i a$ and one has 
\be\label{fstring}
f_{a b} = k_a S \delta_{ab} + \ldots \ ,
\ee
where $k$ is the level of the Kac-Moody algebra.
The $\ldots$ denote perturbative 
and non-perturbative quantum corrections
which are suppressed in the large $S$ (weak coupling)
limit; they play no role in this paper.

The metric of the moduli also simplifies at the 
string tree level. Using the notation 
$\Phi^I = (S,\phi^i)$
with $i=1,\ldots,n_4, I=0,\ldots,n_4$
one has
\be\label{Kstring}
K^{(4)}= -\ln(S+\bar{S}) + \tilde{K}^{(4)}(\phi,\bar\phi) 
+ \ldots\ ,
\ee
where $\tilde{K}^{(4)}(\phi,\bar\phi)$ 
is the tree level K\"ahler potential for
 all moduli but the dilaton.
It is a model dependent 
function and does not enjoy any generic properties.
For Calabi-Yau compactifications with the standard embedding
of the spin connection (so called $(2,2)$ vacua) 
$\tilde{K}^{(4)}$ splits into a sum
\be
\tilde{K}^{(4)} ={K}^{(4)}_{1,1} 
+ {K}^{(4)}_{2,1}\ ,
\label{k4}
\ee
where  ${K}^{(4)}_{1,1} ({K}^{(4)}_{2,1})$
is the  K\"ahler potential for the $(1,1)$-moduli ($(2,1)$-moduli) of $Y_3$.
For future reference we also need to recall that
in the large volume limit of Calabi-Yau compactifications
${K}^{(4)}_{1,1}$ is given by \cite{FS2,CO}
\be\label{k11}
{K}^{(4)}_{1,1} = -\ln[d_{ABC} (t+\bar{t})^A
 (t+\bar{t})^B (t+\bar{t})^C] = 
-\ln[\mbox{Vol}(Y_3)] \ ,
\ee
where $d_{ABC}$ are the (classical) intersection numbers, 
$t^A$ the $(1,1)$-moduli and  Vol($Y_3$) 
is the classical volume 
of the compactification manifold $Y_3$.
For ${K}^{(4)}_{2,1}$ one has instead
\cite{AS,CO}
\be
{K}^{(4)}_{2,1} = -\ln\Big[\int_{Y_3} \Omega\wedge\bar\Omega\, \Big]\ ,
\ee
where $\Omega$ is the unique 
$(3,0)$-form.\footnote{%
${K}^{(4)}_{2,1}$ has a similar expansion as 
${K}^{(4)}_{1,1}$ around the large complex 
structure point \cite{CDGP}.}

Finally, let us note that the couplings of the dilaton
in eqs.\ (\ref{fstring}) and (\ref{Kstring})
are largely determined by the fact that 
$\lambda_4$ organizes the string perturbation theory
and that in perturbation theory 
there is a continuous Peccei-Quinn (PQ) symmetry
$S\to S + i\gamma, \gamma \in \mathbb R$ 
shifting the axion $a$.

\subsection{$D=4,N=1$ supergravity compactified on $S^1$}
Let us reduce the Lagrangian (\ref{eqhet4d}) to $D=3$ on a circle $S^1$.
This does not break any supercharges so that the theory continues
to have 4 real supercharges. In $D=3$ this corresponds to
$N=2$ supersymmetry since the irreducible Majorana spinor
has 2 real components.

For the
$S^1$-reduction to $D=3$ we use the Ansatz \cite{FS}:
\be
g^{(4)}_{mn} = \left( \begin{array}{cc}
                 g^{(3)}_{\mu \nu} + \rr^2 B_{\mu} B_{\nu} & \rr^2 B_{\mu}\\
                 \rr^2 B_{\nu} & \rr^2
                 \end{array}
          \right)\ ,
\qquad  A^{a}_{m} = \left(A^{a}_{\mu} + B_{\mu} \zeta^{a}, \zeta^{a} \right)\ , \label{eqkakl}
\ee
where $\mu,\nu = 0,1,2$ and $\rr$
is the radius (measured in the $D=4$ Einstein 
metric) of the $S^1$. 
The reduction procedure follows closely 
ref.\ \cite{FS} where four-dimensional $N=2$
vacua are considered.
The details for the reduction of 
$N=1$ vacua are shown
in appendix \ref{s1compact} and 
here we only give the results.

In $D=3$ the vector multiplets 
contain the real scalars 
$\zeta^{a}$ in the adjoint representation
of $G$. Thus there is an additional component 
of the moduli space (a Coulomb branch)
spanned by the scalar fields lying in the Cartan subalgebra
of $G$ and at a generic point in this moduli space the
gauge group is  $[U(1)]^{\mbox{r}(G)}$. 
In order to make the notation not too heavy we  continue to label  
the $U(1)$ gauge multiplets by the index $a$ although
in the previous section the same index ran over all
gauge generators;
thus $a=1,\ldots,\mbox{r}(G)$ henceforth.  

In $D=3$ an Abelian vector is (Poincar\'e)
dual to a scalar and 
thus a vector multiplet is dual to a chiral 
multiplet. 
Technically this duality is seen by adding
Lagrange multipliers $C^a, b$ to the 
three-dimensional Lagrangian and eliminating
$A_\mu^a,B_\mu$ by their equations of motion
(see appendix~B).
In this dual picture all supermultiplets are chiral 
and thus their scalar fields have to parametrize a K\"ahler manifold 
\cite{DNT}. It turns out that the K\"ahler 
structure only becomes manifest 
after introducing the coordinates
\beqn\label{DTdef}
D^{a} & \equiv & -f_{ab}(\Phi) \zeta^b + i C^a\ ,  \\
T & \equiv & \rr^2 + i \bb 
+ \frac{1}{2} \R_{ab}^{-1}(\Phi)
D^{a}(D^{b} + \bar{D}^{b})\  .  \nonumber 
\eeqn
Combining all $n_4 + \mbox{r}(G) +2$ scalar fields 
into the coordinate vector  
$Z^\Sigma = (S,\phi^i, D^a,T), \Sigma =0,\ldots,n_4 + \mbox{r}(G) +1$
the three-dimensional Lagrangian can be written as
\be 
{\cal L}^{(3)} =  \frac{1}{2} R^{(3)} 
- G_{\bar\Lambda\Sigma} \pu \bar{Z}^{\bar\Lambda} 
\po Z^{\Sigma} \ ,
\label{l3}
\ee
where $G_{\bar\Lambda\Sigma}$ 
obeys
\beqn
G_{\bar\Lambda\Sigma} &=&\bar\partial_{\bar\Lambda}\partial_{\Sigma}
K^{(3)}_{\mbox{\footnotesize het}} \ , \non
K^{(3)}_{\mbox{\footnotesize het}} &=& 
K^{(4)}(\Phi,\bar{\Phi}) 
- \ln[ T + \bar{T} - \frac{1}{2} (D + \bar{D})^a 
\R^{-1}_{ab} (D + \bar{D})^b ]\ .  
\label{eqhet3dkpot}
\eeqn
 Note that 
the argument of the 
logarithm is given by the square of the 
compactification radius, as can be seen 
from eq.\ (\ref{DTdef}):
\be\label{use}
T + \bar{T} - \frac{1}{2} (D + \bar{D})^a \R^{-1}_{ab} (D + \bar{D})^b 
= 2 \rr^2\ .
\ee

\subsection{The heterotic $D=3$ low energy effective Lagrangian}
\label{secthet3dstr}
So far the reduction did not use any input from string theory
but was merely an `exercise' in supergravity.
Inserting the  string tree level properties 
displayed in eqs.\ (\ref{fstring})
and (\ref{Kstring}) into
$K^{(3)}_{\mbox{\footnotesize het}}$
of eq.\ (\ref{eqhet3dkpot})
yields
\be
K^{(3)}_{\mbox{\footnotesize het}} = \tilde{K}^{(4)} (\phi,\bar\phi)
- \ln \left[(T+\bar{T})(S+\bar{S}) - (D^{a} + \bar{D}^{a})^{2} \right].  \label{k3tree}
\ee
We will see in section \ref{duality} that 
in the dual M-theory vacua we are more naturally 
led to the coordinates $S'= \frac{1}{2}(S+T)$ 
and $T' = \frac{1}{2}(S-T)$ 
and the  K\"ahler 
potential 
\be\label{KK3}
K^{(3)}_{\mbox{\footnotesize 
het}} = \tilde{K}^{(4)} (\phi,\bar{\phi})
- \ln [(S'+\bar{S}')^2 - (T' + \bar{T}')^2 - (D^{a}
 + \bar{D}^{a})^{2} ] \ .
\ee

The next step is to identify the three-dimensional 
dilaton. The relation to the 4-di\-men\-sio\-nal dilaton is as usually
\be\label{3Ddef}
\frac1{\lambda_3^2} = \frac\rs{\lambda_4^2} \ ,
\ee
where $\rs$ is the radius of $S^1$ measured in the 
four-dimensional string frame metric.
In the reduction procedure we used the metric in 
the Einstein frame (eq.\ (\ref{eqkakl}))
which is related to the metric in the string frame
by the Weyl rescaling
$g^{(4)}_{\mbox{\scriptsize E}} = \lambda_4^{-2}
g^{(4)}_{\str}$.
This implies the following relation among the radii 
\be\label{Rrel}
\rs=\rr\lambda_4 \ .
\ee
Combining (\ref{3Ddef}) and (\ref{Rrel}) results in
\be
\lambda_3^2 =  \frac{\lambda_4}\rr  \ .
\label{eqdil}
\ee
Using (\ref{use}) and $\lambda_4^{-2}= \frac12(S+\bar S)$
we also derive
\be
\lambda_3^{-4} = (T+\bar{T})(S+\bar{S}) - (D^{a} + \bar{D}^{a})^{2} \ .
\label{Lambda3}
\ee
The three-dimensional dilaton governs the 
perturbation series in $D=3$, as can be seen by 
reducing the four-dimensional Lagrangian in the 
string frame. The result can be found in appendix 
\ref{s1compact}.
The K\"ahler potential of eq.\ (\ref{k3tree})
is only valid to lowest order 
in $r$ and $\lambda_3$ and gets perturbative 
and non-perturbative corrections.

Finally let us discuss the symmetries of the 
compactified theory.
First of all there is the PQ associated with the
four-dimensional axion 
$a$ discussed in section 2.1.
Furthermore,
there are $\mbox r(G)+1$ Abelian gauge symmetries associated
with the $\mbox r(G)+1$ gauge bosons 
$A_\mu^a,B_\mu$. In the dual Lagrangian
these symmetries appear as continuous 
PQ symmetries  acting on the dual scalars. 
Finally,
the scalars $\zeta^a$ in the three-dimensional vector multiplets
`inherit' another PQ symmetry from the four-dimensional
gauge invariance. So altogether 
(\ref{l3}), (\ref{eqhet3dkpot}) are
 invariant under 
the following $2\mbox r(G)+2$ PQ symmetries 
(with parameters 
$\gamma,\tilde\gamma,\gamma^a,\hat\gamma^a$)
\beqn
a&\to& a + \gamma\ ,\non
\bb &\to& \bb +\tilde \gamma \ , \non
C^a &\to& C^a + \gamma^a \ , \qquad \bb \to \bb +\gamma^a\zeta^a\ ,\\
\zeta^a &\to& \zeta^a + \hat\gamma^a \ , 
\qquad \bb \to \bb +\hat\gamma^a C^a\ .\nonumber
\eeqn
These PQ symmetries are exact in perturbation theory but broken
to discrete subgroups non-perturbatively.
In addition, there is the standard T-duality which acts on the radius of 
the $S^1$ and sends $\rs\to \rs^{-1}$ while keeping $\lambda_3$ fixed. 
Using (\ref{Rrel}) and (\ref{eqdil}) this corresponds to
\be 
\lambda_4 \to \rr^{-1} \ ,\qquad
\rr\to\lambda_4^{-1}
 \ ,
\ee
and also leaves (\ref{k3tree}) invariant.

\section{M-theory compactified on Calabi-Yau fourfolds}
\label{sectm}

M-theory has been suggested as the strong coupling limit
of type IIA string theory \cite{W}. 
Even though  a satisfactory formulation of M-theory has
not been established its low energy limit
is known to be 11-dimensional supergravity.
In this section we perform the compactification of 
this low energy limit on Calabi-Yau fourfolds 
and obtain an effective Lagrangian in $D=3$.

The starting point is the 11-dimensional 
supergravity Lagrangian\footnote{The 
corresponding action is given by 
$S^{(11)} = \int d^{11} x \, 
\sqrt{-g^{(11)}} {\cal L}^{(11)}$. Throughout 
the paper we define the 
Lagrangian without the factor $\sqrt{-g^{(D)}}$.} 
\cite{CJS}:
\be\label{Leleven}
\kappa_{11}^2 {\cal L}^{(11)} = 
\frac{1}{2} R^{(11)} - \frac{1}{4}
  |F_{4}|^{2} - \frac{1}{12} 
(-g^{(11)})^{-1/2} A_{3} \wedge F_{4} \wedge F_{4} \ , 
\ee
where $A_3$ is a three-form, $F_4$ its field strength and 
$g^{(11)}$ the determinant of the 11-dimensional metric 
(more details of the notation used are given in appendix 
\ref{sectnotcon}). 
The Lagrangian (\ref{Leleven}) is the leading order 
contribution in a derivative expansion 
$p/M_{Pl}$ where
$p$ is the characteristic Fourier momentum. 
The next term in this expansion has been deduced from a one-loop 
computation in the type IIA theory \cite{VW} and then 
extrapolated to the 11-dimensional theory. 
It is associated to the sigma-model 
anomaly of the 6-dimensional fivebrane worldvolume 
\cite{DLM}. 
It reads
\beqn 
\delta{\cal L}^{(11)} & \sim & \kappa_{11}^{-2/3}
(-g^{(11)})^{-1/2} A_3 \wedge X_8 \label{X8} \\
\mbox{with} \qquad X_8 & = & \frac{1}{(2 \pi)^4} \left( -\frac{1}{768} 
(\mbox{tr} R^2)^2 + \frac{1}{192} \mbox{tr} R^4 
\right). \nonumber
\eeqn 
(In the following we set $\kappa_{11} = 1$.)

In order to do the reduction of (\ref{Leleven}) let us recall a few facts
about Calabi-Yau fourfolds $Y_4$ \cite{KLRY}. 
They are Ricci-flat K\"ahler manifolds of complex
dimension four with $SU(4)$ holonomy.
The massless modes of the three-dimensional theory are 
determined by the non-trivial harmonic forms 
on $Y_4$ which are the elements of the cohomology
groups $H^{p,q}(Y_4)$. The dimensions $h^{p,q}$ of 
$H^{p,q}$ satisfy
\beqn
&&h^{p,0}=h^{0,p}= h^{4,p}= h^{p,4} = 0 \ , \qquad p = 1,2,3
\ , \nonumber\\
&&h^{0,0}=h^{4,4}=h^{4,0}=h^{0,4}=1\ , \\
&&h^{1,1}=h^{3,3},\quad
h^{1,2}=h^{2,1}=h^{3,2}=h^{2,3},\quad
h^{1,3}=h^{3,1}\ ,\nonumber\\
&&h^{2,2}=2(22+2h^{1,1}+ 2h^{1,3}- h^{1,2})\ ,
\nonumber
\eeqn
leaving $h^{1,1}, h^{1,2}, h^{1,3}$ arbitrary. 
An important further constraint arises from compactification
of the term (\ref{X8}). It induces a potential tadpole
term for the three-form $A_3$ rendering the resulting vacuum
inconsistent \cite{SVW}.
The cofficient of the anomaly is set by the Euler number
$\chi$ of $Y_4$
\be
\int_{Y_4} X_8\ =\ -{\chi\over 24}\ =\ 
- \frac14 (8+h^{1,1}+ h^{1,3}- h^{1,2})\ .
\ee
Thus it
can be avoided  by choosing compactification manifolds
with $\chi=0$. However, the anomaly can also be cancelled
by considering backgrounds with $n$ space-time 
filling
membranes or turning on non-trival $F_4$-flux 
\cite{BB,SVW,DM}.\footnote{A further 
contribution to the 
anomaly related to M5-branes wrapped around
three-cycles in the Calabi-Yau has 
been discussed in \cite{PM}.}
In this case
\be
{\chi\over 24} = n+ \frac1{8\pi^2} \int_{Y_4} F_4\wedge F_4
\ee
has to hold for consistency.
Backgrounds with space-time filling
membranes are known to be dual to heterotic vacua
in non-trivial 5-brane backgrounds \cite{FMW,AC2,AC3}.
Since we are primarily interested in identifying the perturbative 
heterotic string among the fourfold compactifications
we choose to consider $n=0$ in this paper. 
Furthermore, nontrival $F_4$-flux
corresponds in the heterotic dual to discrete twists of the gauge 
bundle \cite{FMW,CD,AC} and can lead to nontrivial 
torsion \cite{DRS,S2}. We will analyze this case in a 
future publication which
leaves us with fourfolds that satisfy $\chi=0$
 as the 
compactification manifolds to consider
in the following.

We now perform a lowest order Kaluza-Klein reduction of the 11-dimensional 
supergravity on Calabi-Yau fourfolds.\footnote{A similar analysis
for IIA compactifications on threefolds can be found in ref.\ \cite{BCF}
while compactification of 11-dimensional 
supergravity on threefolds was considered in  ref.\ \cite{CCDF,AFT}.}
For the 11-dimensional metric we take the Ansatz
\be
{g}^{(11)}_{MN}(x,y) = \left( \begin{array}{cc}
                     {g}^{(3)}_{\mu \nu}(x) & 0 \\
                     0 & {g}^{(8)}_{ab}(x,y)
                     \end{array} \right),
\label{eqprodmetwarp}
\ee
where $x^\mu (\mu = 0,1,2)$ denote the coordinates of  
three-dimensional Minkowski space and $y^a (a = 3, 
\ldots, 10)$ denote the internal Calabi-Yau
coordinates. 
In addition ${g}^{(8)}_{ab}$ depends on a set of moduli $M$
which parameterize the possible deformations of ${g}^{(8)}_{ab}$
compatible with the Calabi-Yau condition. These moduli appear
as massless scalar fields in the three-dimensional effective action
parameterizing flat directions of the effective potential.
In the compactification procedure one chooses
an arbitrary point in this moduli space (this corresponds to choosing
an arbitrary set of vacuum expectation values for the 
scalar fields)
and expands in the
infinitesimal neighborhood of this point
$M= \langle M\rangle + \delta M(x)$.
This induces
\be
{g}^{(8)}_{ab} = \hat{g}^{(8)}_{ab}(\langle M\rangle)
+\delta {g}^{(8)}_{ab}(\langle M\rangle, \delta M(x))\ ,
\ee
where $\hat{g}^{(8)}_{ab}$ is a  background 
metric and 
$\delta {g}^{(8)}_{ab}$ its deformation.
Demanding that $\delta {g}^{(8)}_{ab}$
preserves the Calabi-Yau condition
one can expand it  in terms
of non-trivial harmonic forms on $Y_4$.
It is convenient to introduce complex coordinates 
$\xi_j \, (j=1,\ldots, 4)$ for $Y_4$\ 
defining\footnote{The corresponding 
integration measures are related according to
$d^8 \xi \equiv d^4 \xi \wedge d^4 \bar{\xi} = d^8 y$.} 
$\xi_j = \frac{1}{\sqrt{2}} (y_{2j-1} + i y_{2j})$.
For the deformation of the K\"ahler form  one 
has\footnote{In the following we 
omit the superscript (8) at the internal metric. 
}
\be
i \delta g_{i \bj} 
= \sum_{A=1}^{h^{1,1}} \delta M^{A}(x)\, 
e^{A}_{i \bj}\ ,
\label{eqmetricdeformation}
\ee 
where
$e^{A}$ is an appropriate basis of $H^{1,1}(Y_4)$
and $M^{A}(x)$ are the corresponding (real) moduli.
For the deformations of the complex structure one has
\be
\delta g_{\bi \bj} = 
\sum_{\alpha=1}^{h^{3,1}} 
\delta {Z}^{ \alpha}(x)\,
{b}^{ \alpha}_{\bi \bj}\ , 
\label{eqmetricdeformation2}
\ee  
where ${Z}^{ \alpha}(x)$ are
complex moduli and $b^{\alpha}_{\bi \bj}$ is related to the basis 
$\Phi^{\alpha}$ of $H^{3,1}(Y_4)$
by an appropriate contraction with
the anti-ho\-lo\-mor\-phic 4-form $\bar\Omega$ 
on $Y_4$\  \cite{CO}:
\be
b^{\alpha}_{\bi \bj} = - \frac{1}{3 | \Omega 
|^2} \bar{\Omega}_{\bi}^{klm} 
\Phi^\alpha_{klm \bj}\ ,\qquad
| \Omega |^2 \equiv 
\frac{1}{4!} \Omega_{ijkl} \bar{\Omega}^{ijkl}\ .
\label{bphi}
\ee 
Finally, 
the three-form $A_3$ is expanded in terms of the (1,1)-forms 
$e^A$ and the (complex) (2,1)-forms $\Psi^{I}_{ij\bar{k}}$. 
More precisely 
\be
A_{\mu i\bj} = \sum_{A=1}^{h^{1,1}}
A_{\mu}^{A}(x)\, e^{A}_{i \bj} \ ,
\qquad
A_{ij\bar{k}} = \sum_{I=1}^{h^{2,1}}
N^{I}(x)\, \Psi^{I}_{ij\bar{k}} \ ,
\qquad
A_{\bi\bj k} = \sum_{I=1}^{h^{2,1}}
 \bar{N}^{\bar{J}}(x)\, 
\bar{\Psi}^{\bar{J}}_{\bi\bj k}\ .
\label{eq3formred}
\ee
So altogether 
the compactification leads to 
$h^{1,1}$ vector multiplets $(A_\mu^A,M^A)$
and $h^{2,1}+ h^{3,1}$ chiral multiplets
$(N^I), (Z^\alpha)$.

On the space of $(1,1)$-forms one defines
the metric \cite{S}
\be \label{GAB}
G_{AB} \equiv  \frac{1}{2 {\vol}} 
\int_{Y_4} e^{A} \wedge 
\star e^{B}\ 
= - \frac{1}{2 {\vol}} 
\int_{Y_4} d^{8} \xi
  \sqrt{{g}}
e^A_{i \bj} e^B_{k \bar{m}} 
{g}^{i \bar{m}} {g}^{k \bj},
\ee
where
$\vol$ is the volume of $Y_4$
\be
{\vol} = 
\int_{Y_4} d^{8} \xi \sqrt{{g}} =
 \frac{1}{4!} 
\int_{Y_4} {J} \wedge {J} 
\wedge {J} \wedge {J} = \frac{1}{4!} 
d_{ABCD} {M}^A 
{M}^B {M}^C {M}^D\ .
\label{eqvolumekf}
\ee
$J$ is the K\"ahler form of the Calabi-Yau fourfold
\be
{J} = i {g}_{i \bj} d \xi^i \wedge d \bar{\xi}^{\bj} =
{M} \, \! ^A  e^A
\label{eqkaehlerform}
\ee
and 
$d_{ABCD} = \int_{Y_4} e^A 
\wedge e^B \wedge e^C \wedge e^D$ are the classical 
intersection numbers of $Y_4$. 

On the space of $(3,1)$-forms one defines the metric \cite{AS}
\be \label{Galphabeta}
G_{\alpha \bar{\beta}} \equiv 
\frac{\int_{Y_4}
 \Phi^{\alpha} \wedge \bar{\Phi}^{\bar{\beta}}}
{\int_{Y_4} \Omega \wedge 
\bar{\Omega}} = 
 \frac{1}{4 {\vol}} 
\int_{Y_4} d^{8} \xi
  \sqrt{{g}} b^{\alpha}_{\bj \bar{m}} 
\bar{b}^{\bar{\beta}}_{i k} {g}^{i \bj} 
{g}^{k \bar{m}}  
= \partial_{\alpha} 
\bar\partial_{{\bar \beta}} 
K_{3,1}, 
\ee
which is a K\"ahler metric with K\"ahler potential 
\be
K_{3,1} =
-\ln \Big[ \int_{Y_4} \Omega \wedge 
\bar{\Omega} \Big]. 
\label{K31}
\ee

Finally, on the space of $(2,1)$-forms
we define a metric $G_{I \bar{J}}$ and
intersection numbers $d_{A I \bar{J}}$
\beqn\label{21metric}
G_{I \bar{J}} &\equiv& \frac{1}{4}
\int_{Y_4} \Psi^{I} 
\wedge \star {\Psi}^{{J}}
= \frac{1}{8} 
\int_{Y_4} d^{8} \xi
  \sqrt{{g}} \Psi^I_{ij\bar{k}} 
\bar{\Psi}^{\bar{J}}_{l \bar{m} 
\bar{n}} {g}^{i \bar{m}} {g}^{j \bar{n}} 
{g}^{l \bar{k}}, \\
d_{A I \bar{J}} & \equiv & \int_{Y_4} e^{A} \wedge
\Psi^{I} \wedge \bar{\Psi}^{\bar{J}} = \frac{1}{4} 
\int_{Y_4} d^{8} \xi 
\sqrt{{g}} \epsilon^{ikls} 
\epsilon^{\bj \bar{m} \bar{n} \bar{r}} 
e^A_{i \bj} \Psi^I_{kl\bar{m}} 
\bar{\Psi}^{\bar{J}}_{s \bar{n} \bar{r}}\ . \nonumber 
\eeqn 
The two quantities are related via\footnote{%
This can be checked noticing that 
$
\star \Psi^J = \frac{1}{2} \Big( 
\bar{\Psi}^{\bar{J}}_{i \bj \bar{k}} g^{i \bj} 
g_{l \bar{m}} g_{n \bar{o}} + 
\bar{\Psi}^{\bar{J}}_{l \bar{k} \bar{m}} g_{n \bar{o}}
 \Big) d\xi^l \wedge d\xi^n \wedge 
d\bar{\xi}^{\bar{k}} \wedge 
d\bar{\xi}^{\bar{m}} \wedge 
d\bar{\xi}^{\bar{o}}.
$
The first term corresponds to a term 
$\tilde{\Psi}^{J} \wedge J \wedge J$ with a globally 
defined (0,1)-form $\tilde{\Psi}^{J} = \frac{1}{2} 
\bar{\Psi}^{\bar{J}}_{i \bj \bar{k}} g^{i \bj} 
d\bar{\xi}^{\bar{k}}$, which is closed because of the 
harmonicity of $\bar{\Psi}^{\bar{J}}$ and the fact that 
the K\"ahler form is covariantly constant with respect 
to the Hermitian metric. As $h^{0,1} = 0$ for a 
Calabi-Yau fourfold the first
term is actually absent. 
}
\be
G_{I \bar{J}} = - \frac{i}{4} d_{A I \bar{J}} 
M^A \ , \quad {\rm or}\quad 
d_{A I \bar{J}} 
= 4i\,  {\partial G_{I \bar{J}}\over 
\partial M^A }\ .
\label{eqgij}
\ee
In the following it is convenient to define
a metric independent of $M^A $ as
\be
\Gi_{I \bar{J}} = - \frac{i}{4} c^A  d_{A I \bar{J}} 
\ ,
\label{ghatij}
\ee
where $c^A$ are constant real vectors with no 
vanishing entries but otherwise arbitrary.

It is important to notice is that 
$G_{AB}$ and $d_{ABCD}$ are independent
of the complex structure but
$G_{I \bar{J}}$ and $d_{A I \bar{J}}$
do depend on $Z^\alpha$ and 
$\bar Z^{\bar \alpha}$.
This dependence is not known
generically since it depends on the particular
$Y_4$ under consideration. 
However, as we will show it is not necessary to know
the complex structure dependence of 
$G_{I \bar{J}}$ and $d_{A I \bar{J}}$
explicitly in order
to determine the K\"ahler potential.\footnote{Similarly,
$ K_{3,1}$ is generically known only in terms of
an integral over the holomorphic $(4,0)$-form
$\Omega$ but not necessarily explicitly 
(see (\ref{K31})).}

The  basis $\Psi^I$ of $(2,1)$-forms can locally 
be chosen to 
depend holomorphically on the complex structure
or in other words 
\be
\bar\partial_{\bar Z^{\bar \alpha}}\Psi^I = 0\ ,\qquad
\partial_{Z^\alpha}\Psi^I \neq 0
\ .
\ee
Since $h^{3,0}=h^{0,3}=0$ the derivative 
$\partial_{Z^\alpha}\Psi^I$ can be expanded into
$(1,2)$- and $(2,1)$-forms with
complex-structure dependent
coefficient functions $\sigma$ and $\tau$
\be \label{dzpsi}
 \partial_{Z^\alpha}  \Psi^I = 
\sigma_{\alpha I K}(Z, \bar{Z}) \,
\Psi^K + 
\tau_{\alpha I \bar{L}}(Z, \bar{Z}) \,
\bar{\Psi}^{\bar{L}}\ .
\label{var}
\ee 
Note that $\tau$ is not the complex conjugate
of $\sigma$ but an independent function.
Differentiating (\ref{var}) with respect
to $\bar Z^{\bar \alpha}$ results in  
the following
differential constraints for $\sigma$ and $\tau$
\be
\bar\partial_{\bar Z^{\bar \beta}}\, \sigma_{\alpha I K}
= - \tau_{\alpha I \bar{L}} 
\bar\tau_{\bar\beta \bar{L} K}
\ ,\qquad
\bar\partial_{\bar Z^{\bar \beta}}\,\tau_{\alpha I \bar{K}}
= - \tau_{\alpha I\bar{L}}
\bar\sigma_{\bar \beta \bar{L} \bar K}\ .
\label{dzsigmatau}
\ee
From (\ref{21metric}), (\ref{eqgij}), (\ref{ghatij})  
and (\ref{var}) 
it immediatly follows that the complex structure
dependence of $G_{I \bar{J}}, \Gi_{I \bar{J}},
d_{A I \bar{J}}$ is  constrained
by the differential equations\footnote{%
It would be nice to have a better geometrical
understanding  of eqs.\ (\ref{dzsigmatau}) 
and (\ref{dzgij}). }
\be
\partial_{Z^\alpha} G_{I \bar{J}}
= \sigma_{\alpha I K} G_{K \bar{J}}\ ,\qquad
\partial_{Z^\alpha} \Gi_{I \bar{J}}
= \sigma_{\alpha I K} \Gi_{K \bar{J}}\ ,\qquad
\partial_{Z^\alpha} d_{A I \bar{J}}
= \sigma_{\alpha I K}\, d_{A K \bar{J}}\ .
\label{dzgij}
\ee

The next step is to insert eqs.\
(\ref{eqprodmetwarp})-(\ref{eq3formred})
into (\ref{Leleven}).
The details of this reduction are
presented in appendix \ref{sectkaluzaklein} and 
here we only summarize the results.
The vectors $A_\mu^A$ are again dualized
to scalar fields denoted
by $P^A$. So after dualization the vector multiplet
becomes a chiral multiplet 
with the (real) scalars $(M^A, P^A)$ and
 there are altogether 
$h^{1,1}+h^{1,2}+ h^{1,3}$ chiral multiplets. 
Supersymmetry requires that the Lagrangian of
these chiral multiplets must be expressable 
in the form
\be\label{expect}
{\cal L}^{(3)} =  \frac{1}{2} R^{(3)} 
- G_{\bar\Lambda\Sigma} \pu \bar{Z}^{\bar\Lambda} \po Z^{\Sigma} \ ,
\ee
where $\Lambda,\Sigma = 1, \ldots, 
h^{1,1}+h^{1,2}+ h^{1,3}$
and $G_{\bar\Lambda\Sigma}$ is 
a K\"ahler metric
$G_{\bar\Lambda\Sigma} =\bar\partial_{\bar\Lambda}\partial_{\Sigma}
K^{(3)}_{\mbox{\footnotesize M}}$.
However, it turns out that the scalar fields which appear 
naturally in eqs.\ 
(\ref{eqmetricdeformation})-(\ref{eq3formred})
in the expansion of the harmonic forms
on $Y_4$ 
are not the appropriate K\"ahler coordinates
$Z^{\Sigma}$.
Rather a set of field redefinitions has to
be performed in order to 
 cast the three-dimensional Lagrangian 
 into the form (\ref{expect}).
The proper K\"ahler coordinates are 
$ T^{A}, \nN^I, Z^\alpha$
defined as
\beqn
T^{A} &=& \frac{1}{\sqrt{8}} 
\Big( i P^{A} + \vol G_{AB} M^{B} 
-\frac{i}{8} d_{A M \bar{L}} \Gi^{-1}_{\bar{J} M}
\Gi^{-1}_{\bar{L} I}
\nN^{I} \bar{\nN} \, \! ^{\bar{J}} 
+ \omega_{A I K} \nN^I \nN^K \Big), \label{TA} \\
\nN^I &=& \Gi_{I \bar{J}}
(Z^\alpha, \bar{Z}^{\bar{\alpha}})\, \bar{N}^{\bar{J}}\ , 
\label{Nhat}
\eeqn
while the $Z^\alpha$ are unchanged.
The $\omega_{A I K}$ are functions of $Z^\alpha$ and 
$\bar{Z}^{\bar{\alpha}}$ which have to obey
\be
\bar\partial_{\bar{Z}^{\bar{\alpha}}} \omega_{A I K} = 
-\frac{i}{8} \Gi^{-1}_{\bar{L} I} \Gi^{-1}_{\bar{J} K} 
d_{A M \bar{L}} 
\bar{{\tau}}_{\bar{\alpha} \bar{J} M}\ , 
\label{eqomega}
\ee
but are otherwise unconstrained.
In terms of $ T^{A}, \nN^I, Z^\alpha$
the metric is
K\"ahler with the K\"ahler potential
\be
K^{(3)}_{\mbox{\fn M}}  =  K_{3,1}
 - \ln\Big[\Xi^A \vol G^{-1}_{AB} \Xi^B\Big]\ ,
\label{k3m}
\ee
where
\be 
\Xi^A\equiv 
 \Big( T^{A} + \bar{T}^{A} +
\frac{i}{4 \sqrt{8}}  d_{A M \bar{L}} 
\Gi^{-1}_{\bar{J} M}
\Gi^{-1}_{\bar{L} I}
\nN^{I} \bar{\nN} \, \! ^{\bar{J}} 
- \frac{1}{\sqrt{8}} \omega_{A I K} \nN^I \nN^K 
-\frac{1}{\sqrt{8}} \bar{\omega}_{A \bar{J} \bar{L}} 
\bar{\nN} \, \! ^{\bar{J}} 
\bar{\nN} \, \! ^{\bar{L}} \Big).
\label{Xi} 
\ee
(Note that $\vol$ and ${G}_{AB}$
have to be expressed in terms of the K\"ahler
coordinates.)

Expressed in geometrical terms the argument of the 
logarithm is just the cube of the volume of the 
Calabi-Yau fourfold measured in the M-theory 
metric, which can be checked by inserting (\ref{TA}) into 
(\ref{Xi}) 
\be
K^{(3)}_{\mbox{\fn M}}  = K_{3,1} - 3 \ln \vol.
\label{m3Dk}
\ee
As we will see in the next section the
duality to the heterotic vacua is 
more naturally expressed in terms of
rescaled variables $\tilde{M}^A$
\be
M^A \rightarrow \tilde{M}^A = \vol^{1/2} M^A\ .
\label{matilde}
\ee
Using the rescaled K\"ahler form 
$\tilde{J} = \tm^A \oneone^A$ in (\ref{volab}) and 
(\ref{gab}) one finds 
\be
\tilde{\vol} = \vol^3 \, \! , \qquad
\tilde{G}_{AB} = \vol^{-1} G_{AB} \, , 
\label{gabtilde}
\ee
where
$\tvol$ and 
$\tg_{AB}$ have the same functional dependence 
on $\tm^A$ as $\vol$ and $G_{AB}$ have on $M^A$. 
In terms of the rescaled variables
the K\"ahler potential 
becomes
\be
K^{(3)}_M = K_{3,1} - 
\ln\Big[\Xi^A \tg^{-1}_{AB} \Xi^B\Big]\ 
= K_{3,1} - \ln \tvol\ .
\label{k3mtilde}
\ee

Even though $K^{(3)}_M$ is the sum of 
two terms the moduli space does not
factorize. When expressed in terms of 
the proper K\"ahler coordinates
the second term in eq.\ (\ref{k3mtilde})
does depend on $T^A, \hat N^I, Z^\alpha$
and therefore the metric is not block
diagonal.
However, from eq.~(\ref{Xi}) we learn
that for $\nN^I =0$ the moduli space does 
factorize locally and the K\"ahler potential
becomes  
\beqn\label{Klimit}
K^{(3)}_M &=& K_{3,1}+ K_{1,1} \ ,\nonumber \\
K_{1,1} &=& -
\ln\Big[(T^A+\bar T^A) \tg^{-1}_{AB} (T^B+\bar T^B)\Big]\ ,\\
K_{3,1}&=&
 -\ln 
\Big[\int_{Y_4} \Omega \wedge 
\bar{\Omega}\Big]\ .\nonumber
\eeqn

Finally let us discuss the continuous PQ-symmetries
of the M-theory vacua in the large volume limit.
First of all the scalars $P^A$ which arise from
dualizing the vectors $A_\mu^A$
inherit a PQ-symmetry from gauge invariance.
The K\"ahler potential (\ref{k3m})
is invariant under the shifts
\be
P^A\to P^A + \tilde\gamma^A\ ,
\ee
where $\tilde\gamma^A$ are arbitrary real 
constants.
Secondly,
the $N^I$ arise from expanding the 
three-form in eq.\ (\ref{eq3formred})
and as a consequence 
they `inherit' part of the three-form gauge invariance.
Specifically, the three-dimensional 
Lagrangian is invariant
under the shift
\be
 N^I \Psi^I +  \bar N^{\bar I} 
\bar\Psi^{\bar I}\to 
N^I \Psi^I  +  \bar N^{\bar I} 
\bar\Psi^{\bar I}+ {\rm const.} \hspace{.1cm}.
\ee
In terms of $N^I$ this amounts to 
\be
 N^I\to  N^I + \gamma^I (Z,\bar Z)\ ,\qquad 
\bar N^{\bar I} \to \bar N^{\bar I}   
+ \bar \gamma^{\bar I}  (Z,\bar Z) \ ,
\ee
where the $\gamma^I$ depend on the complex structure
and as a consequence of (\ref{var}) have to
satisfy 
\be
\partial_{Z^\alpha} \gamma^J 
= - \gamma^I \sigma_{\alpha I J}\ ,\qquad
\bar\partial_{\bar Z^{\bar \alpha}}\, \gamma^J 
= - \bar \gamma^{\bar I} \bar\tau_{\bar\alpha \bar{I} J}
\ .
\ee
The redefinition of the $N^I$ in eq.\ (\ref{Nhat})
is precisely such that it renders the 
PQ-symmetry holomorphic
\be
 \hat N^I\to  \hat N^I + \hat \gamma^I (Z)\ ,
\label{holomorphicpq}
\ee
where 
$\hat \gamma^I = \Gi_{I\bar J}\bar \gamma^{\bar J}$
obeys
\be
\bar\partial_{\bar{Z}^{\bar{\alpha}}} \hat{\gamma}^I 
=   0 \ , \qquad 
\partial_{Z^\alpha} \hat{\gamma}^I  
=  \sigma_{\alpha I K} 
\hat{\gamma}^K - \bar{\hat{\gamma}}^{\bar{L}} 
\Gi^{-1}_{\bar{L} K} \tau_{\alpha K \bar{N}} 
\Gi_{I \bar{N}}\ . 
\label{eqgamma}
\ee
The corresponding PQ transformations of the 
fields $T^A$ are
\beqn
T^A & \rightarrow & T^A - \frac{i}{4 \sqrt{8}} 
d_{A M \bar{L}} \Gi^{-1}_{\bar{J} M} 
\Gi^{-1}_{\bar{L} I} \hat{N}^I 
\bar{\hat{\gamma}} \, \! ^{\bar{J}} 
- \frac{i}{8 \sqrt{8}} 
d_{A M \bar{L}} \Gi^{-1}_{\bar{J} M} 
\Gi^{-1}_{\bar{L} I} \hat{\gamma}^I 
\bar{\hat{\gamma}} \, \! ^{\bar{J}} \label{PQTA} \\
& & \mbox{} + \frac{1}{\sqrt{8}} \omega_{A I K} 
\hat{\gamma}^I \hat{N}^K 
+ \frac{1}{2 \sqrt{8}} \omega_{A I K} 
\hat{\gamma}^I \hat{\gamma}^K
 + \frac{1}{\sqrt{8}} \bar{\omega}_{A \bar{I} 
\bar{K}} \bar{\hat{\gamma}} \, \! ^{\bar{I}} 
\bar{\hat{N}} \, \! ^{\bar{K}}
+ \frac{1}{2 \sqrt{8}} \bar{\omega}_{A \bar{I} 
\bar{K}} \bar{\hat{\gamma}} \, \! ^{\bar{I}} 
\bar{\hat{\gamma}} \, \! ^{\bar{K}}. \nonumber
\eeqn
Thus the $P^A$ also transform\footnote{This 
transformation is necessary in order to render 
(\ref{l3dtilde}) invariant under 
(\ref{holomorphicpq}).} 
according to 
\be
P^A \rightarrow P^A - \frac{1}{8} d_{A M \bar{L}} 
\Gi^{-1}_{\bar{J} M} \Gi^{-1}_{\bar{L} I} 
\hat{N}^I \bar{\hat{\gamma}} \, \! ^{\bar{J}}
+ \frac{1}{8} d_{A M \bar{L}} 
\Gi^{-1}_{\bar{J} M} \Gi^{-1}_{\bar{L} I} 
\hat{\gamma}^I \bar{\hat{N}} \, \! ^{\bar{J}}.
\ee
So altogether we have 
$h^{1,1}+2h^{2,1}$ continuous PQ-symmetries
in the large volume limit of the
M-theory compactification.

\section{Duality}
\label{duality}
\subsection{Heterotic -- M-theory duality 
in $D=3$} 
The next step is to investigate the dual relation between
the heterotic and the M-theory vacua.
A first guidance can be obtained by considering 
the 7-dimensional duality 
(M/K3 $\simeq$ Het/T$^{3}$) \cite{W} 
and fibering it over a common complex
2-dimensional base $B_2$.
For large $B_2$ one can apply the adiabatic argument \cite{VW2} 
and conclude that M-theory compactified on a
K3-fibred fourfold should be dual 
to the heterotic string 
compactified on an elliptically fibred threefold. 
The 7-dimensional string coupling constant 
$\lambda_7$ is related to the volume 
$\vol_{K3}$ of the K3 measured in the 
(11-dimensional) M-theory metric
and the respective space-time metrics ($g_7^{\mbox{\fn het}}, 
g_7^M$) differ by a power of $\lambda_7$ \cite{W} 
\be
\lambda_7^{4/3} = \vol_{K3}\ ,  \qquad  g_7^{\mbox{\fn het}} = 
\lambda_7^{4/3} g_7^{M} \ .
\label{eq7d}
\ee 
Using the 
adiabatic argument one derives 
\be
\frac{1}{\lambda_3^2} 
= \frac{\vol_{B_2}^{\mbox{\fn het}}}{\lambda^2_7} 
= \lambda^{2/3}_7\, \vol_{B_2} 
= \vol^{1/2}_{K3}\,  \vol_{B_2},
\label{lambda3}
\ee
where the volume 
of the 
base $B_2$ measured in the heterotic string frame 
metric $\vol_{B_2}^{\mbox{\fn het}}$ 
is related to the volume $\vol_{B_2}$ measured 
in the M-theory metric 
via $\vol_{B_2}^{\mbox{\fn het}} = 
\lambda^{8/3}_7 \vol_{B_2}$ as can bee seen
from eq.\ (\ref{eq7d}).
Eq.\  (\ref{lambda3}) implies that a large volume 
of the Calabi-Yau fourfold corresponds to 
heterotic weak coupling. 
Since the compactification of M-theory 
performed in the previous section is valid 
for large fourfolds 
it is legitimate to compare it
to a weakly coupled heterotic string.
Or in other words the two limits --
large $Y_4$ in M-theory and weak coupling in
the heterotic string -- are mutually compatible.
In particular duality requires that the 
respective  K\"ahler potentials have to agree
in this limit
\be\label{Keq}
K^{(3)}_{\mbox{\fn M}}= K^{(3)}_{\mbox{\footnotesize het}} \ .
\ee
For perturbative heterotic vacua 
where $K^{(3)}_{\mbox{\footnotesize het}}$
is given by eq.\ (\ref{eqhet3dkpot})
the equality between the K\"ahler
potentials is not automatically satisfied
but rather puts a constraint 
on the intersection numbers 
$d_{ABCD}$ or more generally on 
the Calabi-Yau fourfold.
It would be desirable to find those
fourfolds which correspond to a perturbative
heterotic vacuum exactly as was done 
in ref. \cite{AL} for type IIA compactified
on Calabi-Yau threefolds. 
In this paper we only make the first step in this 
direction in that we show that 
for the $(1,1)$ moduli of
K3-fibred fourfolds eq.\ (\ref{Keq}) holds 
in the large base limit.

As the base $B_2$ we take in both theories
a Hirzebruch surface $\mathbb{F}_n$ 
with $n$ even and 
freeze the values of the scalars 
$\nN^{I}, Z^\alpha$ on the M-theory side.
In this case the M-theory K\"ahler potential 
simplifies 
$K^{(3)}_{\mbox{\fn M}} = K_{1,1}$,
where $K_{1,1}$ is given in eq.\ (\ref{Klimit}).
The particular geometry of the Calabi-Yau 
fourfold we are considering affects the form of 
its intersection numbers and thus in view of 
(\ref{gab}) also the form of $\tg_{AB}$. 
But (\ref{gab}) determines $\tg_{AB}$  
in terms of the coordinates $\tm^A$. In order 
to find a map between the heterotic and the 
M-theory variables one has to compare the 
K\"ahler potentials expressed in the  
K\"ahler coordinates. In general it is not 
possible to invert (\ref{TA}) and explicitly
express $\tilde M^A$ (and thus $\tg_{AB}$) 
in terms of the $T^A$. However, as we will see
shortly the explicit relation between
$\tilde M^A$ and $T^A$ can be obtained 
in the large base limit.

In this limit the intersection 
numbers of $Y_4$ enjoy specific properties. 
We therefore need to 
know the group of its divisors\footnote{A similar 
analysis for K3 fibred threefolds can be 
found in section 5.5 of \cite{A}.}, i.e. 
$H_6(Y_4)$. 
Let $\pi: Y_4 \rightarrow B_2$ denote the 
projection map of the K3 fibred fourfold.  
One has three different 
contributions to $H_6(Y_4)$: 
\newcounter{divisor} 
\begin{list}
{\arabic{divisor}.}{\usecounter{divisor} 
\setlength{\rightmargin}{\leftmargin}} 
\item Each divisor $C$ of the base $B_2$ 
contributes a divisor $D$ of $Y_4$ via its pullback 
$D = \pi^{\star} C$.
\item Each monodromy invariant algebraic 2-cycle 
$G_i$ in the generic fibre $F_0$ gives 
a divisor $D_i$ of $Y_4$ if it is transported 
around the base. 
\item When the K3 fibre degenerates over a locus 
of codimension one in the base in such a way that 
the degenerate fibre is reducible, the volumes of 
its components can be varied 
independently. Therefore such reducible ``bad
fibres'' $B_a$  contribute 
further elements to $H_6(Y_4)$. 
\end{list}
 
The intersection of two divisors of the first 
kind can be traced back to the intersection of 
the corresponding divisors in the base:
\be
(D_1 \cdot D_2)_{Y_4} = 
(\pi^{\star} C_1 \cdot \pi^{\star} C_2)
_{Y_4} = 
\pi^{\star} (C_1 \cdot C_2)_{B_2},
\label{eqintersect1}
\ee
which is not a number, but a four-cycle in 
the Calabi-Yau fourfold.  
The subscript indicates the space within 
which we are considering the intersection theory. 
As explained in \cite{MV} $\mathbb{F}_n$ 
(with $n$ even) is topologically the 
product of two $S^2$'s, whose areas are the 
two K\"ahler parameters $U$ and $V$. The 
corresponding divisors $C_U$ and $C_V$ have the 
intersection numbers $(C_U \cdot C_U)_{B_2} 
= (C_V \cdot C_V)_{B_2} = 0$ and 
$(C_U \cdot C_V)_{B_2} = 1$. 
In view of (\ref{eqintersect1}) the 
intersections of the two related divisors 
of the fourfold are either zero or the generic 
fibre, i.e. $(D_U \cdot D_U)_{Y_4}=0$, 
$(D_V \cdot D_V)_{Y_4}=0$ and
$(D_U \cdot D_V)_{Y_4} = F_0$. 
From this fact follows that 
the intersection of three and four 
$D_U,D_V$ automatically vanishes and
that $D_U \cdot D_V$
has no intersection with divisors coming from 
bad fibres $B_a$. Finally, we have  
\be
(D_U \cdot D_V \cdot {D}_i \cdot {D}_j)
_{Y_4} = (G_i \cdot G_j)_{F_0}
\label{eqintersect2},
\ee
where $G_i, G_j$ and ${D}_i, {D}_j$ are 
defined in the second entry of the 
list above. In particular $G_i$ and 
$G_j$ are dual to elements of the (monodromy 
invariant) part of the Picard lattice of the 
generic fibre which 
has signature $(+,-,\ldots,-)$ \cite{A}.
Thus it is 
possible to choose the divisors of $F_0$ in a way 
that the right hand side of (\ref{eqintersect2}) 
is given by $\eta_{i j} = (+, -, \ldots, -)$. 

In the large base limit the (rescaled) 
volume $\tilde{\vol} = 
\frac{1}{4!} d_{ABCD} \tm^A \tm^B \tm^C \tm^D$ is 
dominated by those terms which contain a maximal 
number of base moduli. From what we have just said 
it is clear that the leading contribution is 
\be
\tvol = \frac{1}{2} d_{UVij} \tilde{U} \tilde{V} 
\tm^i \tm^j 
= \frac{1}{2} \tilde{U} \tilde{V} \eta_{ij} \tm^i 
\tm^j,
\label{VlargeB}
\ee
where $\tilde{U} \tilde{V}$ is the 
(rescaled) volume of the base and $\frac{1}{2} 
\eta_{ij} \tm^i \tm^j$ that of the generic fibre. 
Obviously we are in an adiabatic regime, in which 
to leading order the volume is given by the 
product of the base and fibre volumes
$\tvol_{Y_4}=\tvol_{B_2} \tvol_{K3}$.
Using the intersection numbers 
(\ref{VlargeB})  and  (\ref{gab})
we can compute $\tilde{G}_{AB}$
for the moduli
$\tilde U$ and $\tilde V$ of the base and the moduli
$\tilde M^i$ of the generic fibre 
in the large base limit\footnote{At this point
we neglect the possibility of moduli arising
from bad fibres and discuss their 
contribution at the end 
of the section.}
\be
\tilde{G}_{AB} = \left( \begin{array}{ccc}
        \frac{1}{2 \tilde{U}^2} & 
        0 &
        0 \\
        0 &
        \frac{1}{2 \tilde{V}^2} & 
        0 \\
        0 &
        0 &
        -\frac{\eta_{ij}}{\eta_{kl} \tm^k \tm^l} 
        + \frac{2 \eta_{ik} \tm^k \eta_{jl} \tm^l}
{(\eta_{kl} \tm^k \tm^l)^2}
        \end{array}         
        \right).
\label{gablargeB}
\ee
Inserted into (\ref{TA}) using (\ref{gabtilde})
one obtains
\beqn
T^U + \bar{T}^U & = & \frac14 \left( 
\frac{\tilde{V}}{\tilde{U}} \right)^{1/2} 
(\eta_{kl} \tm^k \tm^l)^{1/2} \non
T^V + \bar{T}^V & = & \frac14 \left( 
\frac{\tilde{U}}{\tilde{V}} \right)^{1/2} 
(\eta_{kl} \tm^k \tm^l)^{1/2} \non
T^i + \bar{T}^i & = & \frac12
(\tilde{U} \tilde{V})^{1/2} 
\frac{\eta_{ij} \tm^j}{(\eta_{kl} \tm^k \tm^l)^{1/2}}.
\label{tlargeB}
\eeqn
In connection with (\ref{VlargeB}) this implies 
\be
\tvol = 32 (T^U + \bar{T}^U) 
(T^V + \bar{T}^V) (T^i + \bar{T}^i) 
\eta_{ij} (T^j + \bar{T}^j)\ .
\label{Vttilde}
\ee
Thus in the large $B_2$ limit the volume also
factorizes in the K\"ahler coordinates $T^A$.
The M-theory K\"ahler potential reads in
this limit
\be
K^{(3)}_{\mbox{\fn M}} = -\ln \tvol
= - \ln\Big[(T^U + \bar{T}^U) 
(T^V + \bar{T}^V) 
(T^i + \bar{T}^i) 
\eta_{ij} (T^j + \bar{T}^j)\Big]\ ,
\ee
and can now be compared to the 
$K^{(3)}_{\mbox{\footnotesize het}}$ of 
the heterotic vacua specified in eq.\ (\ref{KK3}).  

On the heterotic side we 
take the large base limit together with 
the large $S$-limit 
(weak four-dimensional coupling).
Thus 
only $S',T',D^a$ and the
(1,1) moduli $U,V$ of the base $B_2$ 
have to be taken into account while
all other (1,1) moduli, 
all (2,1) and all gauge bundle
moduli can be frozen at generic values.
In elliptically fibred threefolds $Y_3$
with the Hirzebruch surface 
$\mathbb{F}_n$ as a base  
the K\"ahler potential of $U$ and $V$ 
simplifies to
$\tilde{K}^{(4)} = -\ln[(U+\bar U)(V+\bar V)]$
\cite{MV,LSTY}.
So altogether we have on the heterotic side
\be\label{Khetl}
K^{(3)}_{\mbox{\footnotesize het}}
 = -\ln [(U+\bar U)(V+\bar V)] 
- \ln [(S'+\bar{S}')^2 - (T' + \bar{T}')^2 - (D^{a}
 + \bar{D}^{a})^{2} ] \ .
\ee
The two  K\"ahler potentials agree if one
identifies
\be\label{map}
(S',T',D^{a}) \leftrightarrow (T^i)\ ,
\qquad (U,V) \leftrightarrow (T^U, T^V)\ .
\ee
We see that 
the moduli which parameterize the
base $B_2$ are identified on both sides
and the $(1,1)$ moduli of the generic K3-fibre
on the M-theory side are identified with
the four-dimensional dilaton, the radius of $S^1$
and the scalars $D^a$ related
to the heterotic vector multiplets.
Note that the 
correspondence (\ref{map})
implies according to eq.\ 
(\ref{Lambda3}) and (\ref{tlargeB}) 
\be
\lambda^{-4}_3 \sim (T^i + \bar{T}^i) 
\eta_{ij} (T^j + \bar{T}^j) 
\sim \tvol_{B_2} = 
\vol_{B_2} \vol,
\label{3ddilaton}
\ee 
which agrees with 
(\ref{lambda3}) obtained in the adiabatic regime.  
Furthermore the number of moduli $T^i$ 
on the M-theory side is bounded by the maximal 
rank of the Picard group of the generic fibre, 
i.e. by $h^{(1,1)}(F_0) = 20$. This is consistent 
with the bound on the heterotic gauge group given 
in (\ref{rGbound}). In fact it is a 
little lower but that  could be 
related to the fact that our analysis is 
based on purely classical geometry. 
(The same issue arises in type IIA vacua
compactified on threefolds \cite{AL}.)

So far we neglected moduli arising from
reducible bad K3 fibres in the fourfold, i.e.
we considered no divisors of the third kind. 
In the context of the
duality between the heterotic string 
compactified
on $K3\times T^2$  and the type IIA 
string on K3-fibred 
$Y_3$ the reducible bad K3 fibres 
were shown to be 
related to non-perturbative physics on the 
heterotic side \cite{AL}.
The same is true for M- and F-theory 
compactifications on K3-fibred
threefolds $Y_3$ \cite{MV,MV2}.
For those K3-fibred fourfolds $Y_4$ which
can be adiabatically obtained as K3-fibred 
threefolds over {\bf $\mathbb{P}^1$}
it follows that the reducible bad K3 fibres
also correspond to non-perturbative
physics on the heterotic side \cite{BM}.
This can be explicitly checked in the
K\"ahler potentials \cite{HL2}.
Finally, the identification of the
$(2,1)$ and $(3,1)$ moduli of $Y_4$
with the elliptic fibre, the $(2,1)$ and the 
bundle moduli of $Y_3$ on the heterotic side
is discussed in 
refs.\ \cite{FMW,BJPS,ACL,CD,BM}.

\subsection{F-theory limit}
Ultimately, one is interested in lifting 
the M-theory/heterotic duality discussed so far
to four space-time dimensions.
This amounts to a simple decompactification 
on the heterotic side while in the dual theory
one has to take the F-theory limit \cite{V}.  
The following discussion of
the F-theory limit is strongly inspired 
by ref.\ \cite{A} where a  
similar decompactification limit
is discussed  
in the context of type IIA/heterotic duality 
in  four 
dimensions. 

Taking the F-theory limit requires that 
the Calabi-Yau fourfold is elliptically fibred.
Here we focus on those fourfolds which are
in addition K3-fibred. Thus the F-theory limit
for this restricted class of fourfolds
requires K3-fibres which themselves are  
elliptically fibred over a base $\mathbb{P}^1$. 
The first step is to blow down any 
rational curves in the generic K3-fibre which take 
the Picard number above its minimal value of 2.
(The two moduli which are always present
arise from the base $\mathbb{P}^1$ and the 
elliptic fibre.) 
The map (\ref{map}) 
implies that on the heterotic side this 
corresponds to freezing the scalars corresponding
to the real part of $D^a$.\footnote{Strictly 
speaking we should first undo the duality transformation
which related the three-dimensional 
vector multiplets
to chiral multiplets since in the F-theory limit
the four-dimensional vector multiplets
(which have no scalar) have to be recovered.}  
After blowing down
the Picard lattice is given by the even self-dual 
lattice $\Gamma_{1,1}$, which we supose to be 
generated by the null vector $v$ and its dual $v^\star$. 
The K\"ahler form of the generic 
K3-fibre is given by 
\be
J = \sqrt{\frac{1}{2\beta}}\, v^\star + 
\sqrt{\frac{\beta}{2}}\, v\ ,\qquad \beta\in 
\mathbb{R}^+\ ,
\label{kaehler}
\ee
where the coefficients have been chosen
such that the volume of the generic fibre in 
M-theory units is of order one. As we are 
only interested in the qualitative 
features of the F-theory limit we do not keep track of 
any constants of order one in the following. 
The class of the 
base of the K3 is given by $v^\star-v$ (having 
self-intersection number $-2$) and the 
class of the elliptic fibre by $v$ 
(with self-intersection number 
$0$). 
The choice $d_{UVij}=\eta_{ij}=(+,-,\ldots,-)$
in our previous discussion
corresponds to choosing
the divisors of the generic K3-fibre 
in such a way that the forms 
representing the cohomology classes of the 
Poincar\' e duals of these divisors are given by 
$A = \frac{1}{\sqrt{2}} (v + v^\star)$ 
and $B = \frac{1}{\sqrt{2}} (v - v^\star)$. 
One verifies that they obey $A.A = 1$, 
$B.B = -1$ and $A.B = 0$.
($A.A \equiv \int_{K3} A\wedge A$, etc.) 
Expanded in this basis the K\"ahler form of the 
generic K3-fibre (\ref{kaehler}) is given by
\be
J = \frac{1}{2} \Big( \frac{1}{\sqrt{\beta}} + 
\sqrt{\beta} \Big) A + \frac{1}{2} \Big( 
\sqrt{\beta} - \frac{1}{\sqrt{\beta}} \Big) B\ .
\label{kaehler2}
\ee
This implies the identification
$M^1 = \frac{1}{2} 
\Big( \frac{1}{\sqrt{\beta}} + 
\sqrt{\beta} \Big)$ and $M^2 =\frac{1}{2} \Big( 
\sqrt{\beta} - \frac{1}{\sqrt{\beta}} \Big)$. 
Inserted into eq.\ (\ref{tlargeB}) 
using (\ref{3ddilaton}) and the fact that the 
volume of the K3 is of order one we derive 
\beqn
T^1 + \bar{T}^1 & \sim & \lambda_3^{-2} 
\left( \sqrt{\beta} + \frac{1}{\sqrt{\beta}} \right) \ ,\non
T^2 + \bar{T}^2 & \sim & - \lambda_3^{-2} 
\left( \sqrt{\beta} - \frac{1}{\sqrt{\beta}} \right)\ . 
\label{tT12}
\eeqn
From the map (\ref{map})
we expect $S' = \frac{1}{2} (S+T) 
\leftrightarrow T^1$ and $T' = \frac{1}{2} (S-T) 
\leftrightarrow T^2$. Thus we obtain 
\beqn
\lambda^{-2}_4 & \sim & S + \bar{S} \sim (T^1 + 
\bar{T}^1) + (T^2 + \bar{T}^2) 
\sim \lambda_3^{-2} 
\frac{1}{\sqrt{\beta}}\ , \label{lambdabeta} \\
r^2 & \sim & T + \bar{T} \sim (T^1 + 
\bar{T}^1) - (T^2 + \bar{T}^2) 
\sim \lambda_3^{-2} 
\sqrt{\beta}.
\label{rbeta}
\eeqn
The volume of the elliptic fibre is given by
\be
J.v = \frac{1}{\sqrt{2 \beta}}
\label{volv}\ ,
\ee
which has to be taken to zero in the 
F-theory limit \cite{V}. 
In view of (\ref{volv}) 
this is equivalent to $\beta \rightarrow \infty$. On the 
heterotic side it corresponds to the expected 
decompactification limit $r \rightarrow \infty$ 
(with $r \lambda_3^2=\lambda_4^2$ fixed) as 
can be seen from (\ref{rbeta}). Finally,
we learn from (\ref{3Ddef}) and (\ref{lambdabeta}) 
that $\sqrt{\beta} \sim r_s$.

\section{Conclusions}

In this paper we derived the K\"ahler potentials 
for the three-dimensional perturbative heterotic 
string (in terms of the four-dimensional one) and 
the K\"ahler potential for the moduli of a 
Calabi-Yau fourfold in the large volume limit 
of M-theory including the moduli coming from the 
three-form potential. We showed that both 
have a common range of validity (weak 
heterotic coupling and large volume of 
the fourfold are compatible with each other) and 
that they do agree for K3-fibred fourfolds 
and elliptic threefolds in 
the large base limit.

It would be desirable to generalize
this analysis and determine all
fourfolds which obey eq.\ (\ref{Keq}).
Moreover,
our analysis was restricted to 
fourfolds with vanishing Euler number 
($\chi=0$) corresponding to the absence
of space-time filling membranes 
and non-trivial four-form flux. 
Incorporating 
such effects requires the presence of a 
non-trivial warp factor in the 
Ansatz for the metric
(\ref{eqprodmetwarp}) \cite{BB}. 

Perturbative quantum corrections
on the heterotic side which correct the 
gauge kinetic function $f_{ab}$ and the
K\"ahler potential can be taken into account
and related to `geometrical corrections'
on the M-theory side. In this manner one might
derive a geometrical version of the heterotic
perturbation expansion \cite{HL2}.
In particular, the holomorphic
$f$-function has to be related to holomorphic
quantities of the fourfold.
Closely related is the issue of 
non-perturbative corrections to the 
superpotential which corresponds to wrapping 
M5-branes over appropriate six-cycles in the 
fourfold \cite{WNPW}.

Finally,
in $D=2$ the duals of the heterotic vacua are 
related to IIA compactified on $Y_4$.
This duality might be used to 
leave the regime of large $Y_4$ and 
compute stringy
($\alpha'$) corrections.
Similarly, issues of 
mirror symmetry in  Calabi-Yau fourfolds
can be studied.

\bigskip\bigskip
\noindent
{\Large {\bf Acknowledgements}}

We would like to thank P. Aspinwall, T. Bauer, 
R. B\"ohm, G.\ Curio, 
H. G\"unther, B. Hunt, W. Lerche, W. Lucht, 
D. L\"ust, P. Mayr, T. Mohaupt, A. Sen, S. Stieberger and 
A. Strominger for useful discussions. 
J.L.\ thanks the ITP, Santa Barbara
for hospitality while part of this work was done.

This work is supported by
GIF (the German--Israeli
Foundation for Scientific Research),
DAAD (the German Academic Exchange Service)
and the Landesgraduierten\-f\"orderung
Sachsen-Anhalt.

\vskip 2cm
\appendix
\noindent
{\Large {\bf Appendix}}

\setcounter{equation}{0}
\setcounter{section}{0}

\section{Notation and conventions}
\label{sectnotcon}

The signature of the space-time
metric is $(-+ \ldots +)$.
The Levi-Civita symbol is defined to transform as a 
tensor, i.e. we have 
\be
\epsilon^{1 \ldots D} =  (\pm g^{(D)})^{-1/2} \qquad 
\mbox{and} \qquad 
\epsilon_{1 \ldots D} =  (\pm g^{(D)})^{1/2},
\label{eqepsilon}
\ee
where the `$-$'-sign has to be used
for space-time indices while 
on the internal Calabi-Yau manifold
the `$+$'-sign is appropriate.
Our conventions for the Riemann curvature tensor 
are
\be
R^{\mu}_{\nu \rho \sigma} = \partial_{\rho} 
\Gamma^{\mu}_{\nu \sigma} - \partial_{\sigma} 
\Gamma^{\mu}_{\nu \rho} + \Gamma^{\omega}_{\nu 
\sigma} \Gamma^{\mu}_{\omega \rho} - 
\Gamma^{\omega}_{\nu \rho} 
\Gamma^{\mu}_{\omega \sigma},
\label{eqcurvature}
\ee
where we use the following definition of the 
Christoffel symbols:
\be
\Gamma^{\mu}_{\nu \rho} = \frac{1}{2} g^{\mu \sigma} 
(\partial_\nu g_{\sigma \rho} + \partial_\rho 
g_{\sigma \nu} - \partial_\sigma g_{\nu \rho}).
\label{eqchrisis}
\ee
The Ricci tensor is defined as
\be
R_{\mu \nu} = R^{\rho}_{\mu \rho \nu}.
\label{eqricci}
\ee
(We are thus using the (+++) conventions of \cite{MTW}.)

Furthermore a $p$-form $A_p$ can be expanded as
\be
A_p = \frac{1}{p!} A_{\mu_1 \ldots \mu_p} 
d x^{\mu_1} \wedge \ldots \wedge d x^{\mu_p},
\label{eqpform}
\ee
which has an obvious generalisation to $(p,q)$-forms:
\be
A_{p,q} = \frac{1}{p! q!} A_{i_1 \ldots i_p 
\bi_1 \ldots \bi_q} d \xi^{i_1} \wedge \ldots \wedge 
d \xi^{i_p} \wedge d \bar{\xi}^{\bi_1} \wedge \ldots 
\wedge d \bar{\xi}^{\bi_q}. 
\label{eqpqform}
\ee
The exterior derivative is defined as
\be
dA_p = \frac{1}{p!} \pu A_{\mu_1 \ldots \mu_p} d x^\mu \wedge
 d x^{\mu_1} \wedge \ldots \wedge d x^{\mu_p}, 
\label{eqextderiv}
\ee
which entails because of 
\be
F_{p+1} = dA_p = \frac{1}{(p+1)!} F_{\mu_1 \ldots \mu_{p+1}} 
d x^{\mu_1} \wedge \ldots \wedge d x^{\mu_{p+1}}
\label{eqfstr}
\ee   
the relation
\be
F_{\mu_1 \ldots \mu_{p+1}} = (p+1) \partial_{[\mu_1} 
A_{\mu_2 \ldots \mu_{p+1}]}.
\label{eqfieldstr}
\ee
With this definition the action for a $p$-form 
potential $A_p$ is given by
\be
-\frac{1}{4} \int d^D x \sqrt{-g^{(D)}} |F_{p+1}|^{2} = 
-\frac{1}{4} \int d^D x \frac{\sqrt{-g^{(D)}}}{(p+1)!} 
F_{\mu_1 \ldots \mu_{p+1}} F^{\mu_1  \ldots \mu_{p+1}}. 
\label{eqpformaction}
\ee 
The Hodge star operator for a $p$-form is defined as
\be
\star A_p = \frac{1}{p!(D-p)!} A_{\mu_1 \ldots \mu_p} 
\mbox{$\epsilon^{\mu_1 \ldots \mu_p}$}_{\nu_{p+1} \ldots 
\nu_D} dx^{\nu_{p+1}} \wedge \ldots \wedge dx^{\nu_D}\ .
\label{eqstar}
\ee
It is generalised to $(p,q)$-forms on a Hermitian 
manifold (where $D$ now denotes its complex dimension) 
by
\beqn
\star A_{p,q} = \frac{(-1)^{(D-p) q}}
{p! q! (D-p)! (D-q)!} 
\bar{A}_{i_1 \ldots i_q \bi_1 \ldots \bi_p} 
\mbox{$\epsilon^{i_1 \ldots i_q}$}_{\bj_{q+1} \ldots 
\bj_D} \mbox{$\epsilon^{\bi_1 \ldots \bi_p}$}_
{j_{p+1} \ldots j_D} \non
\times 
d \xi^{j_{p+1}} \wedge \ldots \wedge d \xi^{j_D} \wedge 
d \bar{\xi}^{\bj_{q+1}} \wedge \ldots \wedge 
d \bar{\xi}^{\bj_{D}},
\label{eqstarcomplex}
\eeqn
where $\bar{A}_{i_1 \ldots i_q \bi_1 \ldots \bi_p} \equiv 
\overline{A_{i_1 \ldots i_p \bi_1 \ldots \bi_q}}$. 
This definition ensures that we have the following 
expression for the scalar product of two $(p,q)$-forms 
on a Hermitian manifold:
\beqn
(A_{p,q},B_{p,q}) &  \equiv & \frac{1}{p! q!} \int 
d^D \xi d^D 
\bar{\xi}\sqrt{g^{(D)}} A_{i_1 \ldots i_p \bi_1 
\ldots \bi_q} \bar{B}^{\bi_1 
\ldots \bi_q i_1 \ldots i_p}  \non
& = &\int A_{p,q} \wedge \star B_{p,q}.  
\label{eqscalarprod}
\eeqn

\section{$S^1$ compactification of $D=4$ 
super\-gravity}
\label{s1compact}

In this appendix we give some details of the 
$S^1$-reduction of four-dimensional supergravity
which follows rather closely ref.\ \cite{FS}. 
Inserting (\ref{eqkakl}) into (\ref{eqhet4d}) and 
performing a Weyl rescaling 
$g^{(3)}_{\mu \nu} \rightarrow \rr^2 
g^{(3)}_{\mu \nu}$ one arrives at 
\beqn
{\cal L}^{(3)} & = & \frac{1}{2} R^{(3)} - 
\frac{1}{\rr^2}\pu \rr \po \rr 
- G_{\bI J} \pu \bar{\Phi}^{\bI} \po \Phi^{J} 
+ \frac{\rr^4}{4}  H_{\mu} H^{\mu} 
- \frac{1}{2\rr^2} \R_{ab} \pu \zeta^{a} \po \zeta^{b}
\non
&+& \frac{\rr^2}{2}  \R_{ab} (F^{a}_{\mu} + \zeta^{a} H_{\mu}) (F^{b \mu} + \zeta^{b} H^{\mu}) 
+ \I_{ab} \pu \zeta^{a} (F^{b \mu} + \zeta^{b} H^{\mu}),
\label{eqhet3d}
\eeqn
where the following abbreviations are used:
\be
H^{\mu}  =  \frac{1}{2} \epsilon^{\mu \nu \rho} 
H_{\nu \rho} = \frac{1}{2} \epsilon^{\mu \nu \rho} 
(\partial_{\nu} B_{\rho} - \partial_{\rho} B_{\nu})  
, \quad 
F^{a \mu}  =  \frac{1}{2} \epsilon^{\mu \nu \rho} 
F^{a}_{\nu \rho} = \frac{1}{2} \epsilon^{\mu \nu \rho}
 (\partial_{\nu} A^{a}_{\rho} - \partial_{\rho} 
A^{a}_{\nu})\ .  
\ee
The vectors can be dualised to scalars by 
adding $\mbox{r}+1$ 
Lagrange multipliers $C^a$ and $\bb$ 
to the Lagrangian
of eq.\ (\ref{eqhet3d})
\be
{\cal L}^{(3)}\to {\cal L}^{(3)} 
-F^{a}_{\mu} \po C^{a} + \frac{1}{2} H_{\mu} \po 
(\bb - \zeta^{a} C^{a}) \ ,  \label{eqlagrm}
\ee
and eliminate the fields $F^{a}_{\mu}$ and $H_{\mu}$ 
via their equations of motion. This results in
\beqn
{\cal L}^{(3)} & = &  \frac{1}{2} R^{(3)} 
- \frac{1}{\rr^2} \pu \rr \po \rr - G_{\bI J} \pu \bar{\Phi}^{\bI} \po 
\Phi^{J}  -\frac{1}{2\rr^2} \R_{ab} \pu 
\zeta^{a} \po \zeta^{b} 
\label{eqhet3ddual} \\
& & \mbox{} \hspace{-1cm} - \frac{1}{4\rr^4}  (\pu \bb + 
\zeta^{a} \stackrel{\leftrightarrow}{\partial}_{\mu} 
C^{a})^{2}  -  \frac{1}{2\rr^2} (\pu C^{a} - \I_{ac} 
\pu \zeta^{c}) \R_{ab}^{-1} (\po C^{b} - \I_{bd} \po 
\zeta^{d}) \ . \nonumber
\eeqn
Expressed in the K\"ahler coordinates (\ref{DTdef})
${\cal L}^{(3)}$ takes the form
\beqn \label{eqhet3dkco} 
{\cal L}^{(3)} & = & \frac{1}{2} R^{(3)} - G_{\bI J} \pu \bar{\Phi}^{\bI} \po \Phi^{J} \\
&-& {\left| \pu T - (D + \bar{D})^{a} \R^{-1}_{ab} \pu D^{b} 
+ \frac{1}{4}(D + \bar{D})^{a} \R^{-1}_{ac} \pu f^{cd} \R^{-1}_{db} (D + \bar{D})^{b} \right|^{2}  \over
\left[T + \bar{T} - \frac{1}{2} (D + \bar{D})^{a} \R^{-1}_{ab} (D + \bar{D})^{b} \right]^{2}}\non
&-&{(\pu D^{a} - \frac{1}{2} \pu f^{ac} \R^{-1}_{cd} (D + \bar{D})^{d}) 
\R^{-1}_{ab} (\po \bar{D}^{b} - \frac{1}{2} \po \bar{f}^{bc} \R^{-1}_{cd} (D + \bar{D})^{d})\over 
\left[T + \bar{T} - \frac{1}{2} (D + \bar{D})^{a} \R^{-1}_{ab} (D + \bar{D})^{b} \right]}. \nonumber
\eeqn
With this form of the Lagrangian one 
verifies eqs.\ (\ref{l3}),
(\ref{eqhet3dkpot}).\footnote{It 
is essential for this to work that $f_{ab}$ 
depends holomorphically on the moduli fields, 
ensuring the identity 
$\partial_{\Phi^I} f_{ab} = 2 
\partial_{\Phi^I} \R_{ab}$.}

For completness let us also give the 
three-dimensional 
Lagrangian of the heterotic string 
in the string frame\footnote{We have used the 
tree level form of the gauge kinetic functions 
(\ref{fstring}) in this formula and 
inserted $\Phi^I = (S,\phi^i)$.}  
\beqn\label{eqhet3dstr}
{\cal L}^{(3)}_{\str} & = & \lambda_3^{-2}
\left( \frac{1}{2} R^{(3)}_{\str} - 
G_{\bi j}^{(4)} \pu \bar{\phi}^{\bi} \po \phi^{j}
+ \frac2{\lambda_3^2} \pu \lambda_3 \po \lambda_3
 - \frac{1}{2\rs^2} \pu \rs \po \rs 
+ \frac{\rs^2}{4}  H_\mu H^\mu \right. \\
&& \left.
 - \frac{1}{2\rs^2} \pu \zeta^a \po \zeta^a 
+ \frac{1}{2} (F_\mu^a + \zeta^a H_\mu)^2 \right)
- \frac{1}{4} \lambda_3^{2} \rs^2 \pu a \po a 
+ a \pu \zeta^a 
(F^{a \mu} + \zeta^a H^\mu)\ . \nonumber
\eeqn
The fact, that we do not get an overall factor 
$\lambda_3^{-2}$ is an artefact of the 
dualisation of the antisymmetric tensor in $D=4$. 
We see that the perturbation series is 
governed by the three-dimensional dilaton 
(\ref{eqdil}) 
which also determines the three-dimensional gauge 
couplings. Notice however, that the gauge coupling 
for the Kaluza-Klein vector $B_\mu$ also depends 
on the fields $\rs$ and $\zeta^a$. 
Dualizing the vectors again yields
\beqn\label{eqhet3dstrdual}
{\cal L}^{(3)}_{\str} & = & \lambda_3^{-2}
\left( \frac{1}{2} R^{(3)}_{\str} - 
G_{\bi j}^{(4)} \pu \bar{\phi}^{\bi} \po \phi^{j}
+ \frac2{\lambda_3^2} \pu \lambda_3 \po \lambda_3
 - \frac{1}{2\rs^2} \pu \rs \po \rs
 - \frac{1}{2\rs^2} 
 \pu \zeta^a \po \zeta^a \right) \non
& & - \lambda_3^{2}
\left( \rs^2
\pu a \po a + \frac{1}{4\rs^2}  (\pu 
b + \zeta^a \stackrel{\leftrightarrow}
{\partial}_{\mu} C^{a})^{2} + \frac{1}{2} 
(\pu C^a - a \pu \zeta^a )^2 \right). 
\eeqn

\section{Dimensional reduction of $D=11$ 
supergravity on Calabi-Yau fourfolds}
\label{sectkaluzaklein}

In this appendix we present some of the technical details 
connected to the dimensional reduction of the 11-dimensional 
supergravity Lagrangian (\ref{Leleven}) on Calabi-Yau 
fourfolds. 
Starting from the Ansatz (\ref{eqprodmetwarp})
one considers infinitesimally
small deformations of the metric, (\ref{eqmetricdeformation}) 
and (\ref{eqmetricdeformation2}),
which preserve the Calabi-Yau condition.
These metric deformations correspond to 
variations of the 
K\"ahler class and the complex structure
 of the Calabi-Yau metric.
To lowest order they
can be deduced as solutions of the 
Lichnerowicz equations, which are derived 
by linearizing the equations for 
Ricci-flatness of the internal metric, i.e. 
$R^{(8)}_{ab}(\hat{g} + \delta g) = 0$.  
Solving these equations shows that 
the metric deformations can (to lowest order) 
be expanded like in (\ref{eqmetricdeformation}) 
and (\ref{eqmetricdeformation2}) \cite{C}. 
The expansions are sufficient as we are 
only interested in the leading terms of the 
Kaluza-Klein reduced Lagrangian.  
The only non-vanishing Christoffel 
symbols apart from $\Gamma_{\mu \nu}^\rho$ and the 
ones with only internal indices are
(in complex coordinates) 
\beqn
\Gamma^j_{\mu i} & = & - \frac{i}{2} g^{\bar{k} j}
\pu M^A \oneone^A_{i \bar{k}} \, + \frac{1}{2} g^{jk} 
\pu \bar{Z}^{\bar{\alpha}} \bb^{\ba}_{ki}\ ,  \non
\Gamma^{\bj}_{\mu i} & = & \frac{1}{2} g^{\bj k} 
\pu \bar{Z}^{\bar{\alpha}} \bb^{\ba}_{ki} \, -
\Big( 
 \frac{i}{2} g^{\bar{k} \bj} \pu M^A 
\oneone^A_{i \bk} \Big) \ ,
\non
\Gamma^i_{\mu \bj} & = & \frac{1}{2} g^{i \bk} 
\pu Z^\alpha b^\alpha_{\bk \bj} \, - \Big( 
\frac{i}{2} g^{ik} \pu M^A \oneone^A_{k \bj} 
\Big) = 
\overline{\Gamma^{\bi}_{\mu j}}\ , \non
\Gamma^{\bi}_{\mu \bj} & = & - \frac{i}{2} 
g^{\bi k} 
\pu M^A \oneone^A_{k \bj} \, + \frac{1}{2} 
g^{\bi \bk} 
\pu Z^\alpha b^\alpha_{\bk \bj} = 
\overline{\Gamma^i_{\mu j}}\ , \non
\Gamma^\mu_{ij} & = & - \frac{1}{2} 
g^{\mu \nu} \partial_{\nu} 
\bz^{\ba} \bb^{\ba}_{ij}\ , \non
\Gamma^\mu_{i \bj} & = & \frac{i}{2} g^{\mu \nu} 
\partial_{\nu} M^A \oneone^A_{i \bj} 
= \Gamma^\mu_{\bj i}\ , \non
\Gamma^\mu_{\bi \bj}  & = & - \frac{1}{2} 
g^{\mu \nu} \partial_{\nu} Z^\alpha b^\alpha_{\bi \bj} 
= \overline{\Gamma^\mu_{ij}}. 
\label{eqchrisis2}
\eeqn
The terms in brackets do not contribute 
to leading 
order in the expansion of $R^{(11)}$. 
The 11-di\-men\-sio\-nal curvature 
scalar splits into
\be
R^{(11)} = 2 g^{i \bj} R^{(11)}_{i \bj} + g^{ij} 
R^{(11)}_{ij} + g^{\bi \bj} R^{(11)}_{\bi \bj} 
+ g^{\mu \nu} R^{(11)}_{\mu \nu},
\label{eq11dcurvaturescalar}
\ee
where we have used the fact, that 
$g^{i \bj} R^{(11)}_{i \bj}$ is real and therefore 
equal to $g^{\bi j} R^{(11)}_{\bi j}$. We have 
\beqn
{R^{(11)}}_{i \bj} & = & {R^{(11) k}}_{i k \bj} 
+ {R^{(11) \bk}}_{i \bk \bj} 
+ {R^{(11) \mu}}_{i \mu \bj}\ ,
\non
{R^{(11)}}_{\mu \nu} & = & 
{R^{(11) k}}_{\mu k \nu} 
+ {R^{(11) \bk}}_{\mu \bk \nu} 
+ {R^{(11) \lambda}}_{\mu \lambda \nu}
\label{eq11driccitensor}
\eeqn
and similar equations for $R^{(11)}_{ij}$ and 
$R^{(11)}_{\bi \bj}$. 
Furthermore certain components of the 11-dimensional 
curvature tensor are related to components of the 
(Ricci-flat) internal and the external 
curvature tensors:
\beqn
{R^{(11) \lambda}}_{\mu \lambda \nu} & = & 
{R^{(3) \lambda}}_{\mu \lambda \nu}\ , \non
{R^{(11) k}}_{i k \bj} & = & {R^{(8) k}}_{i k \bj} + 
\Gamma^\mu_{i \bj} \Gamma^k_{\mu k} -  
\Gamma^\mu_{i k} \Gamma^k_{\mu \bj}, \qquad \mbox{etc. }.
\label{eqcurvaturetensorrelations}
\eeqn
Using these relations and the Christoffel symbols of 
(\ref{eqchrisis2}) one derives to lowest order 
in the moduli
\beqn
&&\frac{1}{2} \int d^{11}x \sqrt{-g^{(11)}} R^{(11)} 
=  \int d^3x \sqrt{-g^{(3)}} \int d^8\xi 
\sqrt{\hat{g}} \Big(\frac{1}{2} R^{(3)} 
- \frac{1}{4} \pu Z^\alpha 
\po \bar{Z}^{\bar{\beta}} 
b^{\alpha}_{\bj \bar{m}}
\bar{b}^{\bar{\beta}}_{ik} \hat{g}^{i \bj} 
\hat{g}^{k \bar{m}}  \non
& &\qquad\qquad\qquad -  \frac{1}{2} \pu M^A 
\po M^B \oneone^A_{i \bj} \oneone^B_{k \bar{m}} 
\hat{g}^{i \bj} 
\hat{g}^{k \bar{m}} + \frac{1}{4} 
\pu M^A \po M^B \oneone^A_{i \bj} \oneone^B_{k \bm} 
\hat{g}^{i \bm} 
\hat{g}^{k \bj} \Big), 
\label{eqrtoquadraticalorder}
\eeqn
where a total derivative has been neglected.

For the reduction of the remaining terms in 
(\ref{Leleven}) we have to expand the three-form 
$A_3$ in terms of the (1,1)-forms $\oneone^A$ 
and (2,1)-forms $\Psi^I$:
\be
A_3 = A^A_\mu dx^\mu \wedge \oneone^A + 
N^I \Psi^I + \bar{N}^{\bar{J}} 
\bar{\Psi}^{\bar{J}}.
\label{3form}
\ee
Using (\ref{dzpsi}) one derives 
\be
F_4 = \frac{1}{2} F^A_{\mu \nu} dx^\mu \wedge 
dx^\nu \wedge \oneone^A + 
D_\mu N^I dx^\mu \wedge \Psi^I 
+ D_\mu \bar{N}^{\bar{J}} dx^\mu \wedge 
\bar{\Psi}^{\bar{J}},
\label{4form}
\ee
where we abbreviated 
\be
D_\mu N^I = \pu N^I + N^K \sigma_{\alpha K I} 
\pu Z^\alpha + \bar{N}^{\bar{L}} 
\bar{\tau}_{\bar{\beta} \bar{L} I} 
\pu \bar{Z}^{\bar{\beta}}, \qquad 
D_\mu \bar{N}^{\bar{J}} = \overline{D_\mu N^J}.
\label{Dmu}
\ee
Inserting (\ref{3form}) and (\ref{4form}) into 
(\ref{Leleven}) one derives to lowest order
\beqn\label{F2red} 
\lefteqn{- \frac{1}{4} \int d^{11} x \sqrt{-g^{(11)}} 
|F_4|^{2}  =}\\
&&
 - \frac{1}{8} 
\int d^3x \sqrt{-g^{(3)}} \int 
d^8 \xi \sqrt{\hat{g}} \Big( 
F^A_{\mu \nu} 
F^{B \mu \nu} \oneone^A_{i \bj} 
\oneone^B_{k \bar{m}} 
\hat{g}^{i \bar{m}} \hat{g}^{k \bj}  
+ D_\mu N^I D^\mu \bar{N}^{\bar{J}} 
\Psi^I_{ij\bar{k}} \bar{\Psi}^{\bar{J}}_{l \bar{m} 
\bar{n}} \hat{g}^{i \bar{m}} \hat{g}^{j \bar{n}} 
\hat{g}^{l \bar{k}} \Big)\nonumber
\eeqn
and\footnote{In order to derive this form of the 
reduced action one has to perform a partial 
integration and use the relation $d_{AI\bar{J}} 
\tau_{\alpha K \bar{J}} = d_{A K \bar{J}} 
\tau_{\alpha I \bar{J}}$, which can be easily derived 
from $\partial_{Z^\alpha} \int \oneone^A \wedge 
\Psi^I \wedge \Psi^K = 0$.} 
\beqn\label{topred}
&&-\frac{1}{12} \int A_{3} 
\wedge F_{4} \wedge F_{4} = \\
&&\qquad -\frac{1}{16} 
\int d^3x \sqrt{-g^{(3)}} 
\epsilon^{\mu \nu \rho} A^A_\mu 
D_{\nu} N^I D_\rho \bar{N}^{\bar{J}}
 \int d^8\xi \sqrt{\hat{g}} 
\epsilon^{ikls} 
\epsilon^{\bj \bar{m} \bar{n} \bar{r}} 
\oneone^A_{i \bj} \Psi^I_{kl\bar{m}} 
\bar{\Psi}^{\bar{J}}_{s \bar{n} \bar{r}}.
\nonumber
\eeqn 

Before we proceed let us define
(in close analogy with \cite{BCF})
\beqn
\vol & \equiv & \frac{1}{4!}
\int_{Y_4} J \wedge J 
\wedge J \wedge J\ , \non
\vol_{A} & \equiv & \frac{1}{4!}
\int_{Y_4} \oneone^A 
\wedge J \wedge J \wedge J, \non
\vol_{AB} & \equiv & \frac{1}{4!}
\int_{Y_4} \oneone^A 
\wedge \oneone^B \wedge J \wedge J 
\label{volab} \\
& = & \frac{1}{12} \int_{Y_4} d^{8} \xi 
\sqrt{g} \oneone^A_{i \bj} \oneone^B_{k \bar{m}} 
g^{i \bar{m}} g^{k \bj}
- \frac{1}{12} \int_{Y_4} d^{8} \xi 
\sqrt{g} \oneone^A_{i \bj} \oneone^B_{k \bar{m}} 
g^{i \bj} g^{k \bar{m}}\ , \nonumber
\eeqn
where $J$ is the K\"ahler form  defined 
in eq.\ (\ref{eqkaehlerform}).
With the help of (\ref{volab})
one derives 
(again in close analogy with  the threefold case
\cite{S})
\beqn
\star \oneone^A & = & \frac{2}{3} 
\frac{\vol_A}{\vol} 
J \wedge J \wedge J 
- \frac{1}{2} \oneone^A \wedge J 
\wedge J\ , \non
G_{AB} & = & - 6 \frac{\vol_{AB}}{\vol} 
+ 8 \frac{\vol_A \vol_B}{\vol^2}
= -\frac{1}{2} \partial_{A} \partial_{B} 
\ln \vol
\label{gab}
\eeqn
and
\be
16 \frac{\vol_A \vol_B}{\vol^2} 
\pu M^A \po M^B 
= \pu \ln \vol \po \ln \vol. 
\label{dlnvol}
\ee

The integrals over the internal coordinates  
in (\ref{eqrtoquadraticalorder}), 
(\ref{F2red}) and (\ref{topred})
can be performed using the definitions (\ref{GAB}), 
(\ref{Galphabeta}) and (\ref{21metric}).\footnote{%
Strictly speaking one has to replace
the Calabi-Yau metric in these definitions
by its background value. This is related to 
the fact that we perform the Kaluza-Klein 
reduction to first non-trivial order
around a fixed but arbitray point in the moduli
space of metrics.
This procedure results in couplings 
in the effective Lagrangian which depend
on the arbitrary background values.
Using field theoretical reasoning this can be
promoted to full non-linear $\sigma$-model
type couplings.}
With the help of (\ref{gab}), (\ref{dlnvol})
and performing a Weyl rescaling 
$g^{(3)}_{\mu \nu} \rightarrow 
\vol^2 g^{(3)}_{\mu \nu}$ one derives 
\beqn\label{M3d}
{\cal L}^{(3)} & = & \frac{1}{2} R^{(3)} - 
\frac{1}{2} \pu \ln \vol \po \ln \vol 
- G_{\alpha \bar{\beta}} \pu Z^\alpha 
\po \bar{Z}^{\bar{\beta}} 
- \vol^{-1} G_{I \bar{J}} 
D_\mu N^I D^\mu \bar{N}^{\bar{J}}  \\
&& -\, \frac{1}{2} G_{AB} 
\pu M^A \po M^B - \frac{1}{4} \vol^2 G_{AB} 
F^A_{\mu \nu} F^{B \mu \nu} - \frac{1}{4} 
 \epsilon^{\mu \nu \rho} 
d_{A I \bar{J}} A^A_{\mu} D_\nu N^I 
D_{\rho} \bar{N}^{\bar{J}}.\nonumber
\eeqn
The three-dimensional vectors are dualised to scalars 
by adding the Lagrange multipliers $P^A$
\be
-F^A_\mu \po P^A \, , \qquad F^{A \rho} = \frac{1}{2} 
\epsilon^{\rho \mu \nu} F^A_{\mu \nu}
\label{lagrangemulti}
\ee
and eliminating the fields $F^A_\mu$ via their 
equations of motion. The result is
\beqn
{\cal L}^{(3)} & = & \frac{1}{2} R^{(3)}  
- G_{\alpha \bar{\beta}} \pu Z^\alpha 
\po \bar{Z}^{\bar{\beta}} - \vol^{-1} G_{I \bar{J}} D_\mu N^I 
D^\mu \bar{N}^{\bar{J}} \non
& & \mbox{} - \frac{1}{2} \pu \ln \vol 
\po \ln \vol - \frac{1}{2} G_{AB} 
\pu M^A \po M^B 
\label{M3ddual} \\
& & \mbox{} -\frac{1}{2 
\vol^2} 
\Big(\pu P^{A} + \frac{1}{8} d_{A K \bar{L}} 
\left( N^{K} D_\mu \bar{N}^{\bar{L}} - 
D_\mu N^K \bar{N}^{\bar{L}} \right) \Big) \non
& & G^{-1}_{AB} \Big(\po P^{B} + \frac{1}{8} 
d_{B I \bar{J}} 
\left( N^{I} D^\mu \bar{N}^{\bar{J}} - 
D^\mu N^I \bar{N}^{\bar{J}} \right) \Big).
\nonumber
\eeqn
In order to 
find the K\"ahler potential for the scalars 
in (\ref{M3ddual}) one introduces the 
coordinates (\ref{TA}), (\ref{Nhat}). 
With this redefinition the derivatives (\ref{Dmu}) 
take the form
\beqn
D_\mu \bar{N}^{\bar{J}} & = & 
\Gi^{-1}_{\bar{J} I} \left[ 
\pu \nN^I + \pu Z^\alpha \left( 
\Gi^{-1}_{\bar{M} K} 
\bar{\nN} \, \! ^{\bar{M}}   
\tau_{\alpha K \bar{L}} \Gi_{I \bar{L}}
- \sigma_{\alpha I K} \nN^K \right) 
\right] \equiv \Gi^{-1}_{\bar{J} I} 
D_\mu \nN^I\ , \non
D_\mu N^I & = & \Gi^{-1}_{\bar{J} I} 
\left[ \pu \bar{\nN} \, \! ^{\bar{J}} + 
\pu \bar{Z}^{\bar{\beta}} \left(
\Gi^{-1}_{\bar{M} K} \nN^K 
\bar{\tau}_{\bar{\beta} \bar{M} L}
\Gi_{L \bar{J}} 
- \bar{\sigma}_{\bar{\beta} \bar{J} \bar{M}} 
\bar{\nN} \, \! ^{\bar{M}} \right)  
\right] \equiv \Gi^{-1}_{\bar{J} I} 
D_\mu \bar{\nN} \, \! ^{\bar{J}}. \non
\label{Dmutilde}
\eeqn
The Lagrangian (\ref{M3ddual}) becomes
\beqn
{\cal L}^{(3)} & = & \frac{1}{2} R^{(3)}  
- G_{\alpha \bar{\beta}} \pu Z^\alpha 
\po \bar{Z}^{\bar{\beta}} - \vol^{-1} G_{I \bar{J}} 
\Gi^{-1}_{\bar{L} I} 
\Gi^{-1}_{\bar{J} M} 
D_\mu \bar{\nN} \, \! ^{\bar{L}} 
D^\mu \nN^M  \non 
& & \mbox{} - \frac{1}{2} \pu \ln \vol \po \ln \vol 
- \frac{1}{2} G_{AB} \pu M^A \po M^B \non
& & \mbox{} -\frac{1}{2 \vol^2} 
\Big(\pu P^{A} + \frac{1}{8} d_{A K \bar{L}} 
\Gi^{-1}_{\bar{N} K} 
\Gi^{-1}_{\bar{L} M} 
\left( \bar{\nN} \, \! ^{\bar{N}} 
D_\mu \nN^M 
- D_\mu \bar{\nN} \, \! ^{\bar{N}} 
\nN^M \right) \Big) \non
& & G^{-1}_{AB} \Big(\po P^{B} + \frac{1}{8} 
d_{B I \bar{J}} \Gi^{-1}_{\bar{P} I} 
\Gi^{-1}_{\bar{J} Q} 
\left( \bar{\nN} \, \! ^{\bar{P}} 
D^\mu \nN^Q 
- D^\mu \bar{\nN} \, \! ^{\bar{P}} 
\nN^Q \right) \Big).
\label{l3dtilde}
\eeqn
Using (\ref{TA})-(\ref{Xi})
one verifies that  ${\cal L}^{(3)}$
given in (\ref{l3dtilde}) coincides with
the Lagrangian of  (\ref{expect}).

\end{document}